\documentclass{tlp}

\usepackage{amsmath}
\usepackage{amsfonts}
\usepackage{amssymb}
\usepackage{hyperref}
\usepackage{url}

\newtheorem{definition}{Definition}
\newtheorem{theorem}{Theorem}
\newtheorem{proposition}{Proposition}
\newtheorem{lemma}{Lemma}
\newtheorem{corollary}{Corollary}
\newtheorem{example}{Example}

\newenvironment{falseTheorem}[2]{\vskip\topsep\noindent{\em #1 #2\\}\ }{\vskip\topsep}
\newenvironment{ftheorem}[1]{\begin{falseTheorem}{Theorem}{#1}}{\end{falseTheorem}}
\newenvironment{fproposition}[1]{\begin{falseTheorem}{Proposition}{#1}}{\end{falseTheorem}}

\newcommand{\nat}{\mathbb{N}}

\renewcommand{\emptyset}{\varnothing}
\renewcommand{\phi}{\varphi}
\newcommand{\toppos}{\mbox{\footnotesize$\Lambda$}} 
\newcommand{\ol}[1]{\overline{#1}} 
\newcommand{\pr}[1]{\mbox{\tt #1}}

\def \tuple#1{\langle #1 \rangle}

\def\defemb#1#2{\expandafter\def\csname #1\endcsname
                              {\relax\ifmmode #2\else\hbox{$#2$}\fi}}
\defemb{cA}{{\cal A}}
\defemb{cB}{{\cal B}}
\defemb{cC}{{\cal C}}
\defemb{cD}{{\cal D}}
\defemb{cE}{{\cal E}}
\defemb{cF}{{\cal F}}
\defemb{cG}{{\cal G}}
\defemb{cH}{{\cal H}}
\defemb{cI}{{\cal I}}
\defemb{cJ}{{\cal J}}
\defemb{cL}{{\cal L}}
\defemb{cM}{{\cal M}}
\defemb{cO}{{\cal O}}
\defemb{cP}{{\cal P}}
\defemb{cR}{{\cal R}}
\defemb{cS}{{\cal S}}
\defemb{cT}{{\cal T}}
\defemb{cU}{{\cal U}}
\defemb{cV}{{\cal V}}
\defemb{cX}{{\cal X}}
\defemb{cZ}{{\cal Z}}

\makeatletter
\newenvironment{prog}
 {\vspace{0.7ex}\par\setlength{\parindent}{10pt}\parskip=0pt
  \obeylines\@vobeyspaces\normalsize\tt}
 {\vspace{0.7ex}\noindent}
\newenvironment{sprog}
 {\vspace{0.7ex}\par\setlength{\parindent}{10pt}\parskip=0pt
  \obeylines\@vobeyspaces\small\tt}
 {\vspace{0.7ex}\noindent}
\makeatother
\newcommand{\startprog}{\begin{prog}}
\newcommand{\stopprog}{\end{prog}\noindent}
\newcommand{\sstartprog}{\begin{sprog}}
\newcommand{\sstopprog}{\end{sprog}\noindent}

\newcommand{\fpr}[1]{\mbox{\footnotesize\tt #1}} 

\newcommand{\Rerased}{{\cal R}_\rho}
\newcommand{\MNRC}{\mbox{\it MNRC\/}}
\makeatletter

\newcommand{\Pos}{{{\cal P}os}}

\defemb{Sfunc}{{\sf S}}
\defemb{Sred}{{\sf red}}
\defemb{Shnf}{{\sf hnf}}
\defemb{Snf}{{\sf nf}}
\defemb{SnfIn}{{\sf nfin}}
\defemb{Seval}{{\sf eval}}
\defemb{SevalN}{{\sf eval_{narr}}}
\defemb{Sempty}{{\sf empty}}
\defemb{REDEX}{{\sf REDEX}}
\defemb{HNF}{{\sf HNF}}
\defemb{NF}{{\sf NF}}
\defemb{NFinf}{{\sf NF}^\omega}

\newcommand{\rarg}[3]{{\it rarg}_{#2}(#3)}

\newcommand{\rarggen}{{\rarg{\cR}{\Sfunc}{f}}}

\newcommand{\rpos}[3]{{\it rpos}_{#2}(#3)}
\newcommand{\rposgen}{{\rpos{\cR}{\Sfunc}{t}}}

\newcommand{\Symbols}{{\cF}}
\newcommand{\CSymbols}{{\cC}}
\newcommand{\DSymbols}{{\cD}}
\newcommand{\Variables}{{\cX}}
\newcommand{\TermsOn}[2]{{\cT(#1,#2)}}

\newcommand{\Terms}{{\TermsOn{\Symbols}{\Variables}}}

\newcommand{\OTerms}{{\cT(\Symbols\cup\{\Omega\},\Variables)}}

\newcommand{\GTerms}{{\cT(\Symbols)}}
\newcommand{\CTerms}{{\cT(\CSymbols,\Variables)}}
\newcommand{\GCTerms}{{\cT(\CSymbols)}}

\newcommand{\TermsRho}{{{\cal T}(\Symbols_\rho,\Variables)}}

\newcommand{\Subst}{{{\it Subst}(\Symbols,\Variables)}}
\newcommand{\GSubst}{{{\it Subst}(\Symbols)}}
\newcommand{\GCSubst}{{{\it Subst}(\cC)}}
\newcommand{\Var}{{\cal V}ar} 

\newcommand{\dom}{{\cD}om}

\newcommand{\restrict}[1]{|_{#1}}

\newcommand{\pwset}{{\cal P}}

\newcommand{\lto}{{\longrightarrow}}

\newcommand{\bsh}{{\backslash}}

\newcommand{\sksT}[1]{{\vec{#1}}}

\newcommand{\Haskell}{{\sf Haskell}}
\newcommand{\Maude}{{\sf Maude}}
\newcommand{\Curry}{{\sf Curry}}

\newcommand{\Pakcs}{{\sf PAKCS}}

\newcommand{\sicstusProlog}{{\sf SICStus Prolog}}
\newcommand{\cime}{{\sc C{\it i}\/ME}}
\newcommand{\aprove}{{\sc AProVE}}

\title[Removing Redundant Arguments Automatically]
{Removing Redundant Arguments Automatically}

\author[M.Alpuente, S. Escobar and S. Lucas]
       {M. Alpuente, S. Escobar and S. Lucas\\
           DSIC, UPV, Camino de Vera s/n,
           E-46022 Valencia, Spain.\\
           \email{\{alpuente,sescobar,slucas\}@dsic.upv.es}}

\submitted{30 December 2003}
\revised{12 May 2005}
\accepted{5 January 2006}

\begin{document}

\label{firstpage}

\maketitle

\begin{abstract}
The application of automatic
transformation processes during the formal development and optimization
of programs can introduce encumbrances in the
generated code that programmers usually (or presumably)
do not write. An example is the introduction of redundant arguments in
the functions defined in the program. Redundancy of a parameter
means that replacing it by any expression does not change the result.
In this work, we
provide methods for the
 analysis and elimination of redundant arguments in term
rewriting systems as a model for
the programs that can be written in more sophisticated languages.
On the basis of the uselessness of redundant arguments, we also propose an
erasure procedure
which may avoid wasteful computations while still preserving
the semantics (under ascertained conditions).
A prototype implementation of these methods has been undertaken,
which demonstrates
the practicality of our approach.
\end{abstract}


\section{Introduction}\label{SecIntroduction}

A number of researchers have noticed that certain processes of
optimization,
transformation, specialization  and reuse
of code often introduce anomalies in the
generated code that programmers usually (or ideally)
do not write  \cite{ASU86,hughes:backwards,LeuschelSorensen:RAF,LS02}.
Examples are redundant arguments in the functions defined
by the program, as well as useless program rules.
The notion of redundant argument
means that replacing it by whatever expression we like, 
the final result does not change;
independently of actual computations.
The following example 
motivates our ideas.

\begin{example}\label{applast}
Consider the following program
that calculates 
the concatenation of two lists of natural numbers
and
the last element of a list, respectively:
\sstartprog
append(nil,y)  = y                 last(x:nil)  = x\nopagebreak
append(x:xs,y) = x:append(xs,y)    last(x:y:ys) = last(y:ys)
\sstopprog
Assume that we  specialize this program
for the call
$\pr{applast(ys,z)} \equiv
	\linebreak
\pr{last(append(ys,z:nil))}$,
which
appends an element
\pr{z} 
at the end of a given list \pr{ys}
and then returns the last element,
\pr{z}, 
of the resulting list;
the example is borrowed from DPPD library of benchmarks \cite{Leu98}
and was also considered in \citeN{LeM95} and \citeN{PP96b} for logic
program specialization.
Commonly, the optimized program
which can be obtained  
 by using an automatic specializer of functional programs \cite{AFJV97,AFV98,AHLV99} is:
\sstartprog
applast(nil,z)  = z                 lastnew(x,nil,z)  = z\nopagebreak
applast(x:xs,z) = lastnew(x,xs,z)   lastnew(x,y:ys,z) = lastnew(y,ys,z)
\sstopprog
The first argument of the function \pr{applast} is
redundant (as well as the
first and second arguments
of the auxiliary function \pr{lastnew}) and would not typically be
written by a programmer who writes  this program by hand.
This program is 
far from
$\{$\pr{applast$'$(ys,z) = lastnew$'$(z)}, \pr{lastnew$'$(z) = z}$\}$,
a more feasible one 
with the same evaluation
semantics, or even the
``optimal'' program---without redundant parameters---$\{$\pr{applast$''$(z) = z}$\}$
which one would ideally expect
(here the rule for
the ``local''  function  \pr{lastnew$'$} is disregarded,
since, after 
optimizing the definition of \pr{applast$'$}, it is not useful anymore).
Note that  standard (post-specialization)
renaming/com\-pression
procedures \cite{AFJV97,Gal93,GS94}
cannot perform this optimization
as they only improve programs where program calls contain dead
functors or multiple
occurrences of the same variable, or the functions are defined by rules
whose rhs's are normalizable.
\end{example}

Therefore, it seems interesting to formalize program analysis
techniques for detecting these kinds of redundancies as well as to
formalize
transformations for eliminating the dead code that appears in the form of
redundant function arguments or useless rules and
which, in some cases, can be safely erased
without jeopardizing correctness.

In this work, 
we investigate the problem of  redundant arguments in Term
Rewriting Systems (TRSs), as a model for the programs
that can be written in more sophisticated equational, functional, or 
functional-logic languages.
We provide a semantic characterization
of redundancy which is parametric w.r.t.\
the observed semantics $\Sfunc$.
After some preliminaries in Section \ref{SecPreliminares}, 
in Section \ref{SecRewSemantics} we consider different (reduction)
semantics $\Sfunc$, including the standard normalization semantics
(typical of pure rewriting)
and the evaluation semantics (closer to functional and equational programming).
In Section \ref{SecRedArgInRewriting} we introduce the notion of
redundancy of an argument w.r.t.\ a semantics $\Sfunc$
and provide some useful properties. 
In Section \ref{SecDecidability} we derive
a decidability result
for the redundancy problem
w.r.t.\ $\Sfunc$
and provide the first effective method for detecting redundancies, which is based on
approximation techniques.
Then, in Section \ref{SecCharRed} we provide a more practical method
to recognize redundancy which allows us to simplify
the general redundancy problem  to the analysis of
the rhs's of the program rules.

At first sight, one could na\"{\i}vely think that redundant arguments
are a straight counterpart of ``needed redex'' positions \cite{HL91}, 
a well-known operational notion in term rewriting, which could be 
easily neutralized by appropriately driving the computation.
Unfortunately, this is not true  as illustrated by the following example.

\begin{example}\label{ExNeededAndRedundancy}
Consider the optimized program of Example \ref{applast} extended with:
\startprog
take(0,xs)      = nil\nopagebreak
take(s(n),x:xs) = x:take(n,xs)
\stopprog
The contraction of redex \pr{take(1,1:2:nil)} at position $1$ in the 
term\footnote{In this paper, 
naturals \pr{1}, \pr{2}, $\ldots$ are often used as shorthand to numbers \pr{s$^n$(0)} where $n=1,2,\ldots$.}
	\linebreak
$t=\pr{applast(take(1,1:2:nil),0)}$\/ 
is {\em needed}  
to normalize the term $t$ to the constructor normal form \pr{0}. 
This means that such redex position (or one of its residuals) must be 
reduced in each rewriting sequence from $t$ to its normal form \pr{0}
(see \cite{HL91}).
However,
the first argument of \pr{applast} is redundant for
normalization, as we showed in Example \ref{applast}, and the program
could be improved by dropping this useless parameter.
Therefore, although needed redexes are an essential piece of the 
computational process which implements the evaluation, from a 
semantic point of view, they can be irrelevant ({\em redundant}).
\end{example}
Since needed redexes must {\em all} be reduced in {\em any} reduction sequence 
leading to a normal form, Example \ref{ExNeededAndRedundancy} shows that
no normalizing reduction strategy is able to dodge the problem by
{\em avoiding} the exploration of the redundant argument. 
Thus, in general, inefficiencies caused by the redundancy of arguments 
cannot be avoided by using rewriting strategies. Therefore, in Section \ref{SecErasingRedArgs}
we formalize an {\em elimination} procedure which gets rid of the redundant 
arguments and provide sufficient conditions for the preservation of
the semantics.
Preliminary experiments in Section \ref{SecExperiments}
indicate that our 
approach is both practical and useful.

An extensive comparison with the related literature
is provided in Section \ref{SecRelatedWork}. 
We summarize some relevant ideas as follows. 
Strictness analysis\footnote{Roughly speaking, a function symbol $f$ 
is strict in its $i$-th argument if any subterm at such argument position
must be completely evaluated during the evaluation of $f$.
In symbols: let $D_1,\ldots,D_k,D$ be  ordered sets with least elements
$\bot_1,\ldots,\bot_k,\bot$ respectively, expressing undefinedness,
a mapping $f:D_1\times\cdots\times D_k\to
D$ is said to be \emph{strict} in its $i$-th argument if
$f(d_1,\ldots,\bot_i,\ldots,d_k)=\bot$ for all
$d_1\in D_1,\ldots,d_k\in D_k$.}
\cite{BHA86,Bur91,Jen91,Myc80,MN92,SPR90,WH87}
can be used to 
determine whether the evaluation of an argument $e_i$ 
within an expression $e=f(e_1,\ldots,e_i,\ldots,e_k)$ 
is ``strictly'' necessary 
to obtain the value of $e$.
The counterpart of this notion
has 
been studied in a number of different analysis techniques
such as 
\emph{dead code analysis} \cite{LS02},
\emph{unneededness analysis} \cite{hughes:backwards}, 
\emph{absence analysis} \cite{CousotCousot94-1}, 
\emph{filtering analysis} \cite{LeuschelSorensen:RAF}, 
or \emph{useless analysis} \cite{WS99}.
Also, similar techniques 
to detect and remove
parts of a program which are computationally irrelevant
have been investigated in the past: 
program 
specialization \cite{AFJV97,AFV98,AHLV99,LeM95,PP94,PP96}, 
slicing \cite{Gouranton98,SD96,RT96,SGM02,Tip95,Weiser84}, 
compile-time garbage collection 
\cite{JM89,PG92,KRS94},
and dead code removal
\cite{BCDG00,Kobayashi00,LS02}. 

In Section \ref{SecNarrowing}, we 
briefly discuss
the detection of redundant arguments 
in functional logic programs mechanized by narrowing.
We conclude in Section \ref{SecConclusions}.
Proofs of all technical results are given in 
\ref{AppProofs}.

This paper is a revised and improved version of \cite{AEL02}.
\nocite{AEL02}

\section{Preliminaries}\label{SecPreliminares}

Term rewriting systems provide an adequate computational model
for functional and equational programming languages 
which allow the definition of functions
by means of patterns, e.g., \Haskell, {\sf Hope}, or
{\sf Miranda} 
\cite{BaaNip_TermRewAllThat_1998,Klo92,PE93}. 
In
the remainder of the paper we follow the standard framework of
term rewriting  for developing our results;
see \cite{BaaNip_TermRewAllThat_1998,Terese03} 
for missing definitions.
In order to simplify our presentation, definitions are given in the one-sorted case;
 the extension to many-sorted signatures is not difficult
\cite{Pad88}, and we comment where they matter  the non-obvious details.

Let $\to\: \subseteq A\times A$ be a binary relation on a set $A$.
We denote 
the inverse of $\to$ by $\leftarrow$,
the symmetric closure
by $\leftrightarrow$,
the transitive closure
by $\to^+$, 
the reflexive and transitive closure by $\to^*$,
and
the reflexive, symmetric and transitive closure by $\leftrightarrow^*$.
We say that $\to$ is {\em confluent} if,
for every $a,b,c\in A$, whenever $a \to^*b$ and
$a \to^*c$, there exists $d\in A$ such that $b \to^*d$ and $c \to^*d$.
We say that $\to$ is 
{\em terminating} (or \emph{well-founded})
iff there is no infinite sequence $a_1~\to~a_2~\to~a_3~\cdots$.

Throughout the paper, $\Variables$ denotes a countable set of
variables $\{\pr{x}, \pr{y}, \pr{w}, \ldots \}$,
and
$\Symbols$ denotes
a finite 
set of function symbols
$\{\pr{f}, \pr{g}, \pr{h}, \ldots \}$,
each one having a fixed arity given by a function
$ar:\Symbols\rightarrow \nat$. 
By $\Terms$ we denote the set of terms and by $\GTerms$ the set of ground
terms, 
i.e., terms without variable occurrences. 
$\Var(t)$ is the set of variables in $t$.
A term is said to be linear
if it has no multiple occurrences of a single variable.
A $k$-tuple $t_1, \ldots, t_k$ of terms is written $\ol{t}$.
The number $k$ of elements of the tuple  $\ol{t}$ will be clarified by the context.

A {\em substitution} is a mapping 
$\sigma:\Variables\to\Terms$ which homomorphically extends to a mapping
$\sigma:\Terms\to\Terms$.
The substitution $\sigma$ is usually different from 
the identity, i.e., $\forall x\in\Variables: id(x)=x$,
for a finite subset $\dom(\sigma)\subseteq \Variables$,
called the \emph{domain} of $\sigma$.
By $\theta\circ\sigma$ we denote the composition of the substitutions $\sigma$ and $\theta$,
i.e., $\theta\circ\sigma(x)=\theta(\sigma(x))$.
Let $\Subst$ denote the set of substitutions and $\GSubst$
be the set of ground substitutions, i.e., substitutions on $\GTerms$.
If $\sigma(t)$ is a ground term,
we call $\sigma$ a \emph{grounding substitution} for $t$.
A {\em unifier} of two terms $t,s$ is a substitution $\sigma$
such that $\sigma(t) = \sigma(s)$ and $\sigma$ is idempotent, i.e., $\sigma\circ\sigma=\sigma$.
A {\em most general unifier} ({\em mgu})
of $t,s$ is a unifier $\sigma$ such that for each unifier $\sigma'$ of
$t,s$ there exists $\theta$ such that $\sigma'=\theta\circ\sigma$.
By $\sigma\restrict{V}$ we denote the restriction of subsitution $\sigma$ to the variables
in $V$. 

Terms are viewed as labelled trees in the usual way. 
Positions $p,q,\ldots$ are defined as sequences of positive natural numbers used to address subterms
of $t$, 
with $\toppos$ the root position (i.e., the empty sequence), 
$p.q$ the position concatenation, and $p < q$ the usual prefix ordering.
Two positions $p,q$ are disjoint, denoted by $p \parallel q$, 
if neither $p < q$, $p > q$, nor $p = q$.
The symbol labeling the root position of $t$ is denoted as $root(t)$.
The subterm at position $p$ of $t$ is denoted as $t|_p$ and  $t[s]_p$ is the
term $t$ with the subterm at position $p$ replaced by $s$.
The restriction of a set of positions $P$ w.r.t. a position $p$ is defined as
$P|_p=\{p' \mid \exists q\in P\wedge q=p.p'\}$,
the concatenation of a position $p$ and a set of positions $P$ is defined as
$p.P=\{p.q \mid q \in P\}$,
and the comparison of a set of positions $P$ w.r.t. a position $p$ is defined as
$p \leq P$ iff $p \leq q$ for each $q\in P$.
By $\Pos_S(t)$ we denote all positions in $t$ with a symbol or variable from 
$S\subseteq \Symbols\cup\Variables$. We use $\Pos_f(t)$ and $\Pos(t)$ as
shorthands for $\Pos_{\{f\}}(t)$ and $\Pos_{\Symbols\cup\Variables}(t)$, respectively.
A context is a term $C$ with
zero or more `holes', i.e., the fresh constant symbol $\Box$.
We usually write simply $C[~]$ to denote an arbitrary context,
clarifying the number and
location of holes `in situ'.
If $C$ is a context and $t$ a term, $C[t]$ denotes the result
of replacing the hole in $C$ by $t$.

A rewrite rule is an ordered pair $(l,r)$, written\footnote{We will use also $l = r$ to differentiate a rule from a rewriting step.} $l\to r$,
with $l,r\in \Terms$, $l\not\in \Variables$ and
$\Var(r)\subseteq \Var(l)$. The left-hand side ({\em lhs}) of
the rule is $l$ and $r$ is the right-hand side ({\em rhs}).
A TRS is a pair $\cR=(\Symbols, R)$ where $R$ is a set of rewrite  rules
and $\Symbols$ is called the signature.
A term $t$ rewrites to $s$ (at position $p$),
written $t \rightarrow_{\cR} s$
(or just $t\to s$), if $t|_p=\sigma(l)$ and $s=t[\sigma(r)]_p$,
for some rule $l\rightarrow r\in R$, $p\in \Pos(t)$ and substitution $\sigma$.
An instance $\sigma(l)$ of the $lhs$ of a rule $l\to r$ is called a redex; 
similarly subterm $t|_p$ in a rewrite step is also called a redex.
A term $t$ without redexes is said a \emph{normal form}. By $\NF_\cR$
we denote the set of finite normal forms w.r.t. $\cR$.
A term $t$ is said a \emph{head-normal form} (or \emph{root-stable}) 
if it cannot be rewritten to a redex.  
By $\HNF_\cR$ we denote the set of head-normal forms w.r.t. $\cR$.

A TRS $\cR$ is \emph{left linear} if
all its lhs's are linear terms.
A TRS $\cR$ is \emph{ground} (resp. \emph{right-ground}) if
all its lhs's and rhs's (resp. only its rhs's) are ground terms.
A TRS $\cR$ is \emph{terminating} (resp. \emph{confluent})
if the relation $\to_{\cR}$ is terminating (resp. confluent).
Two terms $t,s$ are \emph{joinable}, denoted by $t \downarrow s$,
if there exists a term $u$ such that
$t \rightarrow^* u$ and $s \rightarrow^* u$.

Given $\cR=(\Symbols,R)$, we assume $\Symbols$ can be always considered
as the disjoint union  $\Symbols=\CSymbols\uplus\DSymbols$ of
symbols $c\in\CSymbols$, called {\em constructors}, and
symbols $f\in\DSymbols$, called {\em defined functions},
where $\DSymbols=\{f~|~f(\ol{l})\to r\in R\}$
and $\CSymbols=\Symbols-\DSymbols$. Then, $\CTerms$ is the set of
constructor terms. A pattern is a term $f(l_1,\ldots,l_n)$ such
that $f\in\DSymbols$ and $l_1,\ldots,l_n \in\CTerms$.
A constructor system ({\em CS}) is a TRS whose lhs's are patterns.

Two (possibly renamed) rules $l\to r$  and $l'\to r'$ {\em overlap},
if there is a non-variable
position $p\in\Pos_{\Symbols}(l)$ and a most-general unifier $\sigma$ such that
$\sigma(l|_p)=\sigma(l')$. The pair
 $\tuple{\sigma(l)[\sigma(r')]_p,\sigma(r)}$ is called a \emph{critical
 pair}
  and
 is also called an \emph{overlay} if  $p=\toppos$. 
A  critical pair $\tuple{t,s}$ is \emph{trivial} if $t=s$. 
A left-linear TRS without critical pairs is
called {\em orthogonal}.
Note that orthogonality of a TRS $\cR$ implies confluence of $\to_\cR$.
A left-linear TRS where its critical pairs are trivial overlays
is  called {\em almost orthogonal}.

\section{Semantics}\label{SecRewSemantics}

The redundancy of an argument  of a
function $f$ in a TRS $\cR$
depends on the semantics properties of $\cR$
that we are interested in observing.
Our notion of semantics is aimed to couch operational as well as
denotational aspects.

A {\em term semantics} for a signature $\Symbols$
is a mapping
$\Sfunc:\GTerms\to\pwset(\GTerms)$
\cite{Luc01}
which associates a set of terms to a term.
A {\em rewriting semantics} for a TRS $\cR=(\Symbols,R)$ is a term semantics
$\Sfunc$ for $\Symbols$ such that,
for all $t\in\GTerms$ 
and $s\in\Sfunc(t)$, $t\to^*_\cR s$,
i.e., a term semantics where 
the set of terms associated to a term 
is determined only by the program.

The rewriting semantics which is most commonly considered in functional
programming is the set of values (ground
constructor terms) that $\cR$ is able to produce in a finite number
of rewriting steps ($\Seval_\cR(t)=\{s\in\GCTerms \mid t\to^*_\cR s\}$).
Other kinds of rewriting semantics often considered for $\cR$
are, e.g., the set of all possible reducts of a term which are
reached in a finite number of steps
($\Sred_\cR(t)=\{s\in\GTerms \mid t\to^*_\cR s\}$),
the set of such reducts that are ground head-normal forms
($\Shnf_\cR(t)=\Sred_\cR(t)\cap\HNF_\cR$), or
ground normal forms
($\Snf_\cR(t)=\Shnf_\cR(t)\cap\NF_\cR$).
We also consider the (trivial) semantics $\Sempty$ which assigns an empty set
to every term.
We often omit $\cR$ in the notations for rewriting semantics
when it is clear from the context.
Furthermore, 
a rewriting semantics $\Sfunc$ for a TRS $\cR$ is
called
\mbox{\em ($\cR$-)normalized}
if, for all $t\in\GTerms$, $\Sfunc(t)\subseteq\NF_\cR$,
i.e., the semantics associates only normal forms to a term.
$\Seval$ and $\Snf$
are examples of normalized semantics
whereas $\Shnf$ and $\Sred$ are not normalized.

The ordering $\preceq$ between semantics~\cite{Luc01}
provides some interesting properties regarding the
redundancy of arguments.
Given term semantics $\Sfunc$ and $\Sfunc'$ for a signature $\Symbols$, we write
$\Sfunc\preceq\Sfunc'$ if there exists 
$T\subseteq\GTerms$ (called \emph{window set} of $\Sfunc'$ w.r.t. $\Sfunc$)
such that, for all $t\in\GTerms$, $\Sfunc(t)=\Sfunc'(t)\cap T$.
Note that, then,
we have
$\Sempty\preceq\Seval_\cR\preceq\Snf_\cR\preceq\Shnf_\cR\preceq\Sred_\cR$.

Given a rewriting semantics $\Sfunc$,
it is interesting to determine whether $\Sfunc$
provides non-trivial information for every input expression.
Let $\cR$ be a TRS and $\Sfunc$ be a rewriting semantics for $\cR$, we say
that $\cR$ is {\em $\Sfunc$-defined} if for all $t\in\GTerms$,
$\Sfunc(t)\neq\emptyset$~\cite{Luc01}.
$\Sfunc$-definedness is monotone w.r.t.\ $\preceq$: if
$\Sfunc\preceq\Sfunc'$
and $\cR$ is $\Sfunc$-defined, $\cR$ is also $\Sfunc'$-defined.

$\Sfunc$-definedness
has already been studied in the literature
for different semantics~\cite{Luc01}.
In concrete, 
$\Snf$-defined TRSs are known as normalizing 
TRSs (i.e., every term has a normal form \cite{BaaNip_TermRewAllThat_1998}) and 
$\Seval$-definedness is related to termination and 
the standard notion of {\em completely defined\/} (CD) TRSs; see
\cite{KapNarZha_OnSuffCompAndRelPropOfTRSs_AI87,Kounalis_CompDataTypeSpec_EUROCAL85}.
Roughly speaking,
a defined function symbol is completely defined if it does not occur in any
ground term in normal form, that is to say that functions
are reducible on all ground terms (of appropriate sort).
A TRS $\cR$ is {\em completely defined} if each defined symbol of the
signature is completely defined.
In one-sorted theories,
completely defined programs occur only rarely. However, they are
common  when using types, and each function is defined for all constructors of
its argument types.

Let $\cR$ be a normalizing 
 and completely defined TRS; then,
$\cR$ is $\Seval_\cR$-defined.
Being completely defined is sensitive to extra constant symbols in
the signature, and so is redundancy. Thus, we are not concerned with
modularity in this work. 

{From} now on,
we formulate the notion of a redundant argument 
and provide 
some useful properties and detection techniques.

\section{Redundant Arguments}\label{SecRedArgInRewriting}

Roughly speaking, a redundant argument of a function $f$
is an argument $t_{i}$ which we do not need to consider in order to
compute the semantics of any
call containing a subterm $f(t_1,\ldots,t_k)$.

\begin{definition}[Redundancy of an argument]
\label{context}
Let
$\Sfunc$ be a term semantics for a signature $\Symbols$, $f\in\Symbols$, and
$i\in\{1,\ldots,ar(f)\}$.
The $i$-th argument of $f$ is {\em redundant
w.r.t.\ $\Sfunc$\/} if,
for all contexts $C[\;]$ and for all $t,s\in \GTerms$ such that $root(t)=f$,
 $\Sfunc(C[t])=\Sfunc(C[t[s]_i])$.
\end{definition}

We denote by
$\rarggen$ the set of redundant arguments of a symbol
$f\in\Symbols$ w.r.t.\ a semantics $\Sfunc$ for $\Symbols$. Note that
every argument of every symbol is redundant w.r.t.\ $\Sempty$.
The following result shows that redundancy is antimonotone
with regard to the ordering $\preceq$ on semantics.

\begin{theorem}[Antimonotonicity of redundancy]\label{TheoRedundancyIsAntiMonotone}
Let $\Sfunc,\Sfunc'$ be term semantics
for a signature $\Symbols$. If $\Sfunc\preceq\Sfunc'$, then, for all
$f\in\Symbols$, $\rarg{\Symbols}{\Sfunc'}{f}\subseteq\rarg{\Symbols}{\Sfunc}{f}$.
\end{theorem}

The following result guarantees that constructor symbols have no
redundant arguments for usual non-trivial semantics, 
which agrees with the common understanding of constructor terms as
completely meaningful pieces of information.

\begin{proposition}[Non-redundancy of constructors]\label{PropThereIsNoRedArgForConstructors}
Let $\cR$ be a TRS such that $|\GCTerms|>1$, and
consider a rewriting semantics $\Sfunc$  such that $\Seval_{\cR}\preceq\Sfunc$.
Then, for all $c\in\CSymbols$,
$\rarg{\cR}{\Sfunc}{c}=\emptyset$.
\end{proposition}

\noindent
For many-sorted signatures, we would require that 
$|\GCTerms_{\tau}|>1$ for the sort $\tau$
of an argument of a constructor symbol $c$.
In the following section, we consider several aspects about decidability of
the redundancy of an argument.

\section{Decidability Issues}\label{SecDecidability}

In general, the redundancy of an argument is undecidable.
However, we are able to provide 
a decidability result about redundancy w.r.t. all the non-trivial semantics
considered in this paper. 
In this section, 
for a signature $\Symbols$, term semantics
$\Sfunc$ for $\Symbols$, $f\in\Symbols$, and $i\in\{1,\ldots,ar(f)\}$,
by ``redundancy w.r.t.\ $\Sfunc$''
we mean the redundancy of the $i$-th argument of
$f$ w.r.t.\ $\Sfunc$.

We follow the ``(W)S$k$S approach'' to decide a given property $P$,
which is based on ascertaining the conditions for expressing $P$ in a
{\em decidable} logic, namely {\em the (weak) second-order monadic logic
with $k$ successors \emph{(W)S$k$S}}; 
see \cite{Tho90}.
The following theorem by Rabin is the key element for our results 
in this section.

\begin{theorem}[\citeNP{Rabin69}]\label{TheorRabin} 
The (weak) monadic second-order theory of $k$ successor functions
\emph{(W)S$k$S} is decidable.
\end{theorem}

First, we recall some basic definitions about
the WS$k$S logic; see e.g., \cite{Tho90}.
{\em Terms} of the WS$k$S logic are formed out of individual
variables $x,y,z,\ldots$, the empty string $\toppos$, and right
concatenation with $1,\ldots,k$. Atomic formulas are equations
between terms, inequations $w < w'$ between terms, or expressions
$w \in X$ where $w$ is a term and $X$ is a (second-order)
variable. Formulas are built from atomic formulas using the
logical connectives $\wedge, \vee, \Rightarrow, \neg, \ldots$ and
the quantifiers $\exists, \forall$ of both individual and
second-order variables. Individual variables are interpreted as
elements of $\{1,\ldots,k\}^{\ast}$ and second-order variables as
finite subsets of $\{1,\ldots,k\}^{\ast}$. Equality is the string
equality and inequality is the strict prefix ordering.
Finite union and intersection, as well as
inclusion and equality of sets, are definable in WS$k$S in an
obvious way.

Let us relate TRSs and WS$k$S logic.
Given a finite signature $\Symbols$, let $k$ be the maximal arity of
all the function symbols in $\Symbols$ and $n$ be the cardinality of $\Symbols$.
A term $t$ is represented in WS$k$S using $n+1$ set
variables $X$ and $X_f, f\in\Symbols$, which are denoted
by $\sksT{X}$ in the following.
$X$ will be the set of all positions of $t$,
and $X_f$ will be the set of positions
that are labeled with the corresponding function symbol.
The following WS$k$S formula
expresses that $\sksT{X}$ encodes a term in $\GTerms$
\cite{Com00,DM97}:
\[\centering
Term_\Symbols(\sksT{X}) \stackrel{def}{=}
\begin{array}[t]{@{}l@{}}
X = \bigcup_{i=1}^{n} X_{f_i} 
\wedge \bigwedge_{i\neq j} (X_{f_i} \cap X_{f_j} = \emptyset) \\
\wedge ~\forall x \in X \; \forall y<x \,(y\in X) \\
\wedge ~
\bigwedge_{f\in\Symbols}(
   \forall x\in X_f: \begin{array}[t]{@{}l}
    \bigwedge_{l=1}^{ar(f)} (x.l \in X) \wedge 
    \bigwedge_{l=ar(f)+1}^{k} (x.l \not\in X))\end{array}
\end{array}
\]
If $Term_\Symbols(\sksT{T})$ holds, then we let $t_{\sksT{T}}$ define
the term in $\GTerms$ which is
uniquely determined by $\Pos(t)=T$ and $root(t|_p)=f$ if $p\in T_f$
for all $p\in T$. A subset of ground terms $L \subseteq \GTerms$
is called WS$k$S {\em definable} if
there exists a WS$k$S formula $\Phi$
with free variables $\sksT{T}$
such that
$L=\{t_{\sksT{T}}\mid Term_\Symbols(\sksT{T}) \wedge \Phi(\sksT{T})\}$.

An arbitrary term semantics $\Sfunc$ can be encoded as a relation
$\cS$ between terms:
$\cS=\{(t,s)\mid t\in\GTerms\wedge s\in\Sfunc(t)\}$. Hence, we
say that semantics $\Sfunc$ is WS$k$S definable if there exists a WS$k$S
formula $\Phi$ with free variables $\sksT{T}$ and $\sksT{S}$ such that
$(t_\sksT{T},s_\sksT{S})\in\cS\Leftrightarrow
Term_\Symbols(\sksT{T})\wedge Term_\Symbols(\sksT{S})\wedge
\Phi(\sksT{T},\sksT{S})$.

\begin{theorem}[Decidability of redundancy]
\label{TheoDeciRed}
Let $\Sfunc$ be a term semantics for a signature $\Symbols$.
If $\Sfunc$ is WS$k$S definable, then
redundancy w.r.t.\ $\Sfunc$\/
is decidable.
\end{theorem}

\noindent
The following result shows that decidability of redundancy is
{\em antimonotone} with regard to the ordering $\preceq$ on semantics.

\begin{proposition}\label{TheoDecRedIsAntiMonotone}
Let $\Sfunc,\Sfunc'$ be term semantics for
a signature $\Symbols$. If $\Sfunc \preceq \Sfunc'$, $\Sfunc'$ is WS$k$S definable,
and there exists a window set $T \subseteq \GTerms$
of $\Sfunc'$ w.r.t.\ $\Sfunc$ which
is WS$k$S definable,
then $\Sfunc$ is WS$k$S definable.
\end{proposition}

In 
\cite{DHLT87,DHLT90},
ground (finite) tree transducers (GTT for short) were introduced
to recognize the rewrite relation $\to^*_\cR$ in
(left-linear and right-)ground TRSs.
Since GTT-recognizable relations are definable in
WS$k$S~\cite{Com00},
the semantics $\Sred$
is also WS$k$S definable, hence the redundancy w.r.t.\ $\Sred$ is decidable.
Now, the following result shows that the window set $\HNF_{\cR}$
is WS$k$S definable; this is useful for proving that semantics $\Shnf$
is also WS$k$S definable.

\begin{theorem}\label{TheoNHFDefinable}
The set $\HNF_{\cR}$ of a finite left-linear, right-ground TRS $\cR$ is
WS$k$S definable.
\end{theorem}

Then, the following theorem provides the first 
decidability result w.r.t. all the non-trivial semantics
considered in this paper. 

\begin{theorem}[Decidability for semantics $\Sred_\cR$, $\Shnf_\cR$, $\Snf_\cR$, and $\Seval_\cR$]\label{CoroDecidRed}
For a left-linear, right-ground TRS $\cR$ over a finite signature
$\Symbols$, the redundancy w.r.t.\ semantics
$\Sred_\cR$, $\Shnf_\cR$, $\Snf_\cR$, and $\Seval_\cR$
 is decidable.
\end{theorem}

This result recalls the decidability of other related properties of
TRSs, such as confluence, joinability, and reachability
problems (for left-linear, right-ground TRSs) \cite{DHLT90,Oya90}.
For instance, the
confluence problem was shown to be undecidable for right-ground TRSs,
while it is decidable for ground TRSs and also for
left-linear and right-ground TRSs \cite{DHLT90}.
Note that we cannot weaken in our approach the requirement
 of right-groundness in
Theorem \ref{CoroDecidRed}
to
the more general conditions of shallowness \cite{Com00} or growingness
\cite{Jacq_DecidApproxTRS_RTA96} as the induced rewrite relations
are not expressible in
the logic WS$k$S
that we use to decide the property 
\cite{DM97}.

In the following section we provide the first redundancy detection method,
which
(sufficiently) ensures that an argument
is redundant in a given TRS. 

\subsection{Approximations of Redundancy}\label{approxs}

Whenever a property is undecidable or costly to decide, we use
approximations.
A notion of approximation (for TRSs) that
has been proven useful for approximating interesting properties in term
rewriting (namely neededness of redexes for normalization) is the
following  \cite{DM97,%
Jacq_DecidApproxTRS_RTA96}:
Given TRSs $\cR$ and $\cR'$ (possibly with extra variables)
over the same signature,
$\cR'$ {\em approximates} $\cR$ if
$\to^*_\cR\subseteq\to^*_{\cR'}$
and $\NF_\cR=\NF_{\cR'}$. An approximation of TRSs
is a mapping $\alpha$ from TRSs to TRSs with the property that
TRS $\alpha(\cR)$ approximates TRS $\cR$ \cite{DM97}.
We write $\cR_\alpha$ instead of $\alpha(\cR)$ to denote the
approximation of $\cR$ according to $\alpha$.
{\em Strong}, {\em nv} \cite{DM97}, {\em shallow} \cite{Com00}, 
and {\em growing} \cite{Jacq_DecidApproxTRS_RTA96}
are examples
of such approximations of TRSs. In all these approximations,
the rhs's of the rules are modified in different ways. For
instance, given a TRS $\cR$, $\cR_{nv}$ is obtained by replacing {\em
all} variables in the rhs by new, different variables that do {\em
not} occur in the lhs; this is possible
since the framework deals with extra variables.

In order to approximate redundancy, we need to use a new symbol $\Omega$
to represent all ground terms (in particular, to be used at the
argument position which is tested for redundancy).
Inspired by \cite{DM97,Oya86}, we
define our notion of  approximation as follows.
Let $\cR$ be a TRS over a signature $\Symbols$ and $\cR'$ be a TRS
 over the signature $\Symbols\cup\{\Omega\}$, where $\Omega$ is a new constant
symbol defined by the rules $\{\Omega\to f(\ol{\Omega})\mid
f\in\Symbols\}$. We extend the approximation notion of
\cite{DM97,%
Jacq_DecidApproxTRS_RTA96} naturally to TRSs over signatures
$\Symbols$ and $\Symbols\cup\{\Omega\}$, where $\Omega$ is a special
symbol that potentially expresses any term.
Note that 
we consider the normalization semantics
only for ground terms.
Thus, we say that $\cR'$
approximates $\cR$ (but notice that, now, $\cR'$ is a TRS on
$\Symbols\cup\{\Omega\}$) if
$\to^*_\cR\cap\:(\GTerms\times\GTerms )\subseteq\to^*_{\cR'}\cap\:(\GTerms\times\GTerms )$
and $\NF_\cR=\NF_{\cR'}$.
Note that,
$\to^*_\cR\subseteq(\Terms\times\Terms )$ whereas
$\to^*_{\cR'}\subseteq(\OTerms\times\OTerms)$;
however,
by definition
of $\cR'$, $\NF_{\cR'}\subseteq\GTerms$.

The following notation is auxiliary.

\begin{definition}[$\Sfunc$-determinacy w.r.t.\ $f$ and $i$]
Given a symbol $f\in\Symbols$ and
an argument $i\in\{1,\ldots,ar(f)\}$, we say that the semantics
$\Sfunc$ is \emph{determined} w.r.t.\ $f$ and $i$ if
for every context $C[\:]$ and $t\in\GTerms$ such that $root(t)=f$,
then $|\Sfunc(C[t[\Omega]_i])|\leq 1$; 
where $|A|$ stands for the cardinality of the set $A$.
\end{definition}

The following theorem provides a sufficient condition for redundancy
which is the basis of our decidable approximations of redundancy.

\begin{theorem}[Approximation of redundancy]%
\label{TheoApproximationOfRedundancyOfEval}
Let $\cR=(\Symbols,R)$ be a TRS, $\cR'$ be an approximation of $\cR$,
 $f\in\Symbols$, $i\in\{1,\ldots,ar(f)\}$,
and $\Sfunc\in\{\Seval,\Snf\}$. If $\cR$ is $\Sfunc_\cR$-defined
and $\Sfunc_{\cR'}$ is determined w.r.t.\ $f$ and $i$,
then $i\in\rarg{\cR}{\Sfunc_\cR}{f}$.
\end{theorem}

It is an open problem whether redundancy is decidable for terminating
TRSs. Nevertheless, Theorem \ref{TheoApproximationOfRedundancyOfEval}
ensures that redundancy w.r.t.\ $\Snf$ is approximable
for terminating TRSs, since any terminating TRS $\cR$ is
$\Snf_\cR$-defined.
The following theorem ensures that WS$k$S definability of a semantics
entails the possibility of 
guaranteeing decidability of 
a given approximation.

\begin{theorem}[Decidability of $\Sfunc$-determinacy w.r.t.\ $f$ and $i$]%
\label{TheoDecidRedViaDeterminism}
Let $\Sfunc$ be a term semantics for a signature
$\Symbols\cup\{\Omega\}$.
If\/ $\Sfunc$ is WS$k$S definable, then
it is decidable whether
$\Sfunc$ is determined w.r.t.\ $f$ and $i$.
\end{theorem}

Remember that the 
semantics $\Seval_\cR$ and $\Snf_\cR$ are WS$k$S definable for left-linear,
right ground TRSs over finite signatures. 
This suggests us to use the
following approximation of left-linear right-ground TRSs.
Given $\cR=(\Symbols,R)$,
we define $\cR_{rg}=(\Symbols\cup\{\Omega\},R_{rg})$ as follows:
\[R_{rg}=\{l\to r_\Omega\mid l\to r\in R\}\cup\{\Omega\to f(\ol{\Omega})\mid
f\in\Symbols\}\]
where $t_\Omega$ is the term $t$ with all variables replaced by
$\Omega$. It is straightforward to see that $rg$ is an
approximation of TRSs.
The following theorem ensures that 
$\Sfunc$-determinacy w.r.t.\ $f$ and $i$
is decidable for an approximation $\cR_{rg}$ of a TRS $\cR$
and semantics $\Sfunc\in\{\Snf_{\cR_{rg}},\Seval_{\cR_{rg}}\}$.

\begin{theorem}\label{TheoDecidRedViaDeterminismForRG}
Let $\cR$ be a left-linear TRS, $\cR_{rg}$ be the approximation $rg$ of $\cR$,
 $f\in\Symbols$, $i\in\{1,\ldots,ar(f)\}$,
and $\Sfunc\in\{\Seval_{\cR_{rg}},\Snf_{\cR_{rg}}\}$.
It is decidable whether
$\Sfunc$ is determined w.r.t.\ $f$ and $i$.
\end{theorem}

By Theorems \ref{TheoApproximationOfRedundancyOfEval}
and
\ref{TheoDecidRedViaDeterminismForRG},
redundancy of an argument w.r.t.\ $\Snf_\cR$ (and $\Seval_\cR$)
is {\em effectively} approximable by using $rg$.

\begin{corollary}[Approximation of redundancy for $\cR_{rg}$]%
\label{CoroApproximationOfRedundancyOfEval}
Let $\cR=(\Symbols,R)$ be a left-linear TRS,
 $f\in\Symbols$, $i\in\{1,\ldots,ar(f)\}$,
and $\Sfunc\in\{\Seval,\Snf\}$. 
If $\cR$ is $\Sfunc_\cR$-defined
and $\Sfunc_{\cR_{rg}}$ is determined w.r.t.\ $f$ and $i$,
then $i\in\rarg{\cR}{\Sfunc_\cR}{f}$.
\end{corollary}

\begin{example}\label{ExUsingApproxForRedundancy}
Consider the left-linear TRS $\cR$ 
\startprog
f(x,0) = 0            f(0,s(y)) = s(0)      f(s(x),s(y)) = g(x,y)\nopagebreak
g(x,y) = f(x,s(y))
\stopprog
Note that $\cR$ is terminating, hence $\Snf_\cR$-defined.
Approximation $\cR_{rg}$ is:
\startprog
f(x,0) = 0            f(0,s(y)) = s(0)      f(s(x),s(y)) = g($\Omega$,$\Omega$)\nopagebreak
g(x,y) = f($\Omega$,s($\Omega$))   $\;\Omega$ = f($\Omega$,$\Omega$)           $\Omega$ = g($\Omega$,$\Omega$)         
$\Omega$ = s($\Omega$)             $\;\Omega$ = 0
\stopprog
It is not difficult to see that
$\Snf_{\cR_{rg}}$ is determined w.r.t.\ \pr{f} and $1$ 
whereas is not determined w.r.t.\ \pr{f} and $2$.
It is possible to construct an automaton which tests those conditions,
see e.g., \cite{TW68} for more details, thus making it
automatically provable.
By Theorem \ref{TheoApproximationOfRedundancyOfEval},
this means that $1\in\rarg{\cR}{\Snf_\cR}{\pr{f}}$.
\end{example}
The approximation $rg$ is similar
to $nv$ of \cite{DM97}, 
that replaces every variable in rhs's by fresh ones. 
However, including the new symbol $\Omega$ in the 
rhs's of the approximated program is essential for  our
development
since the semantics of the program obtained by the approximation $nv$
is not expressible in the logic WS$k$S.

In the following section, we address the
redundancy analysis from a complementary perspective.
Rather than going more deeply in the decidability issues,
we are interested in ascertaining conditions 
which
(sufficiently) ensure that an argument
is redundant in a given TRS. In order to address this problem, we
investigate redundancy of positions.

\section{Redundancy of positions}\label{SecCharRed}

When considering a particular (possibly non-ground) function call, we can
observe a
more general
 notion of redundancy which allows us to consider arbitrary (deeper) positions
 within the call.

\begin{definition}[$p$-prefix-equal terms]
We say that two terms $t,s\in\Terms$ are
$p$-{\em prefix-equal\/}, with $p\in\Pos(t)\cap\Pos(s)$ if,
 for all occurrences $w$ with $w<p$,
$t|_{w}$ and $s|_{w}$ have the same symbol at the root.
\end{definition}

\begin{definition}[Redundant position]
\label{DefStrongRedPos}
Let $\Sfunc$ be a term semantics for a signature $\Symbols$ and
$t\in\Terms$.
The position  $p\in\Pos(t)$
is {\em redundant} in $t$ w.r.t.\ $\Sfunc$ if,
for all $t',s \in \GTerms$
such that $t$ and $t'$ are $p$-prefix-equal,
$\Sfunc(t')=\Sfunc(t'[s]_p)$.
\end{definition}

\noindent
We denote by $\rposgen$ the set of redundant positions of a term
$t$ w.r.t.\ a semantics $\Sfunc$.

Note that the previous definition
cannot be simplified by getting rid of $t'$ and
simply requiring that 
for all $s \in \GTerms$,
$\Sfunc(t)=\Sfunc(t[s]_p)$,
mimicking Definition \ref{context}.
The reason is that positions in a term cannot be analyzed independently 
for redundancy if we want our notion of redundancy of positions to be truly compositional,
as the following example shows.

\begin{example}\label{no-compos}
Let us consider the TRS $\cR$:
\startprog
f(a,a) = a      f(a,b) = a      f(b,a) = a     f(b,b) = b
\stopprog
Given the term $t=\pr{f(a,a)}$, for all terms $s\in\GTerms$,
$\Seval_\cR(t[s]_1)=\Seval_\cR(t)$ and 
	\linebreak
$\Seval_\cR(t[s]_2)=\Seval_\cR(t)$.
However, $\Seval_\cR(t[\pr{b}]_1[\pr{b}]_2)\neq\Seval_\cR(t)$.
Indeed, 
	\linebreak
$1,2 \not\in \rpos{}{\Seval_{\cR}}{t}$.
\end{example}

In the following, we extend Theorem \ref{TheoRedundancyIsAntiMonotone}
and Proposition \ref{PropThereIsNoRedArgForConstructors} (which 
concern redundant arguments of function symbols)
to redundant positions of terms.

\begin{theorem}[Antimonotonicity of redundancy of a position]\label{TheoRedundancyPositionIsAntiMonotone}
Let $\Sfunc,\Sfunc'$ be term semantics
for a signature $\Symbols$. If $\Sfunc\preceq\Sfunc'$, then, for all
$t\in\Terms$, $\rpos{}{\Sfunc'}{t}\subseteq\rpos{}{\Sfunc}{t}$.
\end{theorem}

\begin{proposition}[Non-redundancy of constructor positions]\label{PropNoRposCTerm}
Let $\cR$ be a TRS such that $|\GCTerms|>1$, and
 $\Sfunc$ be a rewriting semantics such that $\Seval_{\cR}\preceq\Sfunc$.
Then, for all $t\in\CTerms$,
$\rpos{\cR}{\Sfunc}{t}=\emptyset$.
\end{proposition}

The following result states that the positions of a term which
are below the indices addressing the redundant arguments of any function symbol
occurring in $t$ are redundant. 

\begin{proposition}\label{PropRedArgAndRedPos}
Let $\Sfunc$ be a term semantics for a signature $\Symbols$, $t\in\Terms$,
 $p\in\Pos(t)$, $f\in\DSymbols$.
For all positions $q,p'$ and $i\in\rarggen$ such that
$p=q.i.p'$ and $root(t|_q)=f$,
$p\in\rposgen$ holds.
\end{proposition}

In the following, we provide some general criteria for
ensuring redundancy of arguments on the basis  of the (redundancy
of some) positions in the rhs's
of program rules, specifically the positions of the rhs's
where the arguments of the functions defined in the lhs's `propagate' to.
Theorems \ref{TheoRedundancyIsAntiMonotone} 
and \ref{TheoRedundancyPositionIsAntiMonotone}
 say that the more
restrictive a semantics is, the more redundancies  there are for the
arguments of function symbols.
According to our hierarchy of semantics (by $\preceq$),
$\Seval$ seems to be the most fruitful
semantics for analyzing redundant arguments.
In the following, we focus on the problem of characterizing the
 redundant arguments
w.r.t.\ $\Seval$.

\subsection[Characterizing Redundancy: the Variable Case]
           {Using Redundant Positions for Characterizing Redundancy: \linebreak the Variable Case}
\label{SecUsingRedPosCharRedVarCase}

In this section,
we focus on the problem of characterizing the
redundant arguments
w.r.t.\ $\Seval_\cR$ by studying the redundancy  w.r.t.\  $\Seval_\cR$ of some
positions in the rhs's of program rules.
The following definition is useful to detect 
whether the variables of the $i$-th argument in a lhs of symbol $f$ 
propagate to positions in the rhs under the same $i$-th argument of symbol $f$.

\begin{definition}[$(f,i)$-redundant variable]\label{DefRedVar}
Let $f\in\DSymbols$, $i\in\{1,\ldots,ar(f)\}$,
and $t \in \Terms$.
The variable $x \in \Variables$ is {\em $(f,i)$-redundant in $t$}
if it occurs only at positions $p\in\Pos_x(t)$ 
which 
(i) are redundant w.r.t. $\Seval_\cR$ in $t$,
i.e., $p\in\rpos{}{\Seval_\cR}{t}$,
or 
(ii) they appear inside the $i$-th parameter of $f$-rooted subterms of $t$,
i.e., $\exists q$ such that $q.i \leq p$ and $root(t|_q)=f$.
\end{definition}

\noindent
Note that variables which do not occur in a term $t$ are trivially
$(f,i)$-redundant in $t$ for any $f\in\Symbols$ and $i\in\{1,\ldots,ar(f)\}$.

\begin{example}\label{applast:RedVar}
Consider the rules for symbol \pr{lastnew} 
in Example \ref{applast}:
\startprog
lastnew(x,nil,z) = z       lastnew(x,y:ys,z) = lastnew(y,ys,z)
\stopprog
Variable \pr{x} is $(\pr{lastnew},1)$-redundant in rhs's $r_1=\pr{z}$ and $r_2=\pr{lastnew(y,ys,z)}$,
since it does not appear in them.
Variable \pr{ys} is $(\pr{lastnew},2)$-redundant in rhs $r_2$,
since it appears under the second argument of symbol \pr{lastnew}.
\end{example}

Now, we are able to provide the second effective method
to determine redundant arguments based on the $(f,i)$-redundant variables
occurring in rhs's.
In order to prove Theorem
\ref{TheorRedAndVarInRhs} below,
we introduce some auxiliary definitions and lemmata.

Given a TRS $\cR=(\Symbols,R)$, we
write $\cR_f$ to denote the TRS $\cR_f=(\Symbols,\{l\to r\in R\mid
root(l)=f\})$ which contains the set of rules defining $f\in\DSymbols$.
The following definition provides the set of positions of 
the $i$-th parameter of $f$ symbols in $t$.
\begin{definition}
Let $f\in\Symbols$, $i\in\{1,\ldots,ar(f)\}$,
and $t\in\Terms$.
We define 
$Pos_{f,i}(t)=\{q.i\in\Pos(t) \mid root(t|_q)=f\}$.
\end{definition}

Let $\ol{t}=t_1,\ldots,t_n$ be a sequence of terms,
$P=p_1,\ldots,p_n$ be a sequence of positions of another term $s$,
and
$P'=p'_1,\ldots,p'_m$ be a subsequence of $P$
(i.e., $m < n$ and $\exists \mu:\{1,\ldots,m\} \to \{1,\ldots,n\}$ 
such that $p'_i=p_{\mu(i)}$ and $i<i' \Rightarrow \mu(i)<\mu(i')$),
we denote $\ol{t}|_{P'}=t'_1,\ldots,t'_m$
such that $t'_i=t_{\mu(i)}$.
The following result is auxiliary 
and proves 
that the same constructor term is obtained by rewriting
when we replace the set of subterms
at $\Seval_\cR$-redundant and $Pos_{f,i}$ positions in a term
by an arbitrary set of terms.

\begin{proposition}\label{PropRedAndVarInRhs}
Let
$\cR$ be a left-linear CS,
$f \in \DSymbols$, and $i\in\{1,\ldots,ar(f)\}$.
Let $t\in\GTerms$,
$P\subseteq Pos_{f,i}(t)\cup\rpos{}{\Seval_\cR}{t}$
be a set of disjoint positions,
and $\ol{s}\in\GTerms$.
Let $t \to^* \delta$ for some $\delta\in\GCTerms$.
If,
for all $l\to r\in\cR_f$, $l|_i$ is a variable
which is $(f,i)$-redundant
in $r$,  then 
$t[\ol{s}]_P \to^* \delta$.
\end{proposition}

Now, we provide the second effective method to detect
redundancy.

\begin{theorem}[Detecting redundancy: the Variable Case]\label{TheorRedAndVarInRhs}
Let $\cR$ be a left-linear CS. Let
$f \in \DSymbols$ and $i\in\{1,\ldots,ar(f)\}$.
If,
for all $l\to r\in\cR_f$, $l|_i$ is a variable
which is $(f,i)$-redundant
in $r$,  then $i\in\rarg{\cR}{\Seval_\cR}{f}$.
\end{theorem}

\begin{example}\label{Exabogus}
A standard example in the literature on
{\em useless variable elimination} (UVE)---a popular technique for removing dead variables,
see \cite{WS99,Kobayashi00}---is the following program\footnote{The original example
uses natural \pr{100} as stopping criteria for the third argument,
while we simplify here to natural
\pr{1} in order to code it only with two rules.}
with constructor symbols $\CSymbols=\{\pr{0},\pr{s}\}$
and variables \pr{a}, \pr{bogus}, and \pr{j}:%
\startprog
loop(a,bogus,0) = loop(s(a),s(bogus),s(0))\nopagebreak
loop(a,bogus,s(j)) = a
\stopprog
Here it is clear that the second argument does not contribute to the
value of the computation.
By Theorem~\ref{TheorRedAndVarInRhs},
the second argument of \pr{loop}
is redundant w.r.t.\ $\Seval_\cR$.
\end{example}

The restriction to left-linear rules in Theorem \ref{TheorRedAndVarInRhs} above is 
not strictly necessary; however, in most practical cases
the redundancy of the argument of symbol $f$  cannot be analyzed independently 
when we consider repeated variables in left-hand sides, as witnessed by the following example.

\begin{example}
Consider the TRS $\cR$:
\startprog
f(x,x) = a
\stopprog
where \pr{f} and \pr{a} are the only function symbols in the signature. 
Since every 
ground term $t$ rewrites to $\pr{a}$ (this can be easily proved by 
structural induction), both arguments of \pr{f} are redundant w.r.t.\ 
$\Seval_\cR$. 
However, if we add a new constant symbol \pr{b}, then no argument of 
\pr{f} is redundant anymore.
\end{example}
The following example demonstrates that the restriction to
constructor systems in Theorem \ref{TheorRedAndVarInRhs}  is also necessary. 

\begin{example} 
Consider the following non-constructor TRS $\cR$ where $\cC=\{\pr{a},\pr{b}\}$:
\startprog
f(a,x) = g(f(b,x))    g(f(b,x)) = x
\stopprog
Then,  the second argument of \pr{f(a,x)} in the lhs of the first rule
is a variable which, in the corresponding rhs of the rule,
 occurs within the second argument of a subterm rooted by $f$, namely \pr{f(b,x)}.
Hence, by Theorem~\ref{TheorRedAndVarInRhs} we would have that
$2\in\rarg{\cR}{\Seval_\cR}{\pr{f}}$.
However,
$\Seval_\cR(\pr{f(a,a)})=\{\pr{a}\}\neq\{\pr{b}\}=\Seval_\cR(\pr{f(a,b)})$,
which contradicts $2\in\rarg{\cR}{\Seval_\cR}{\pr{f}}$.
\end{example}

Moreover, the extension of this result to 
the normalization semantics $\Snf$ is not possible, as shown in the following example.

\begin{example} 
Consider the TRS $\cR$ where $\CSymbols=\{\pr{a},\pr{b}\}$:
\startprog
f(a,x) = a
\stopprog
This TRS satisfies the conditions of Theorem~\ref{TheorRedAndVarInRhs} and
then $2\in\rarg{\cR}{\Seval_\cR}{\pr{f}}$.
In concrete, we have that, for all $s$, $ \Seval_\cR(\pr{f(b,$s$)})=\emptyset$.
However, $\Snf_\cR(\pr{f(b,a)})=\{\pr{f(b,a)}\}\neq\{\pr{f(b,b)}\}=\Snf_\cR(\pr{f(b,b)})$.
\end{example}

Now, we are able to detect some redundancies in Example \ref{applast}.

\begin{example}\label{applast:lastnew:1}
Let us revisit the following rules from the CS $\cR$ of Example \ref{applast}:
\startprog
lastnew(x,nil,z) = z         lastnew(x,y:ys,z) = lastnew(y,ys,z)
\stopprog
Using Theorem
\ref{TheorRedAndVarInRhs}, 
we are able to conclude that the first argument of
function \pr{lastnew} is (trivially) redundant w.r.t.\ $\Seval_\cR$,
since, in every lhs, the first parameter of \pr{lastnew} 
is a variable that is $(\pr{lastnew},1)$-redundant in the respective rhs.
\end{example}

Unfortunately, Theorem \ref{TheorRedAndVarInRhs} does not suffice to
prove that the {\em second} argument of \pr{lastnew}
is redundant w.r.t.\ $\Seval_\cR$,
and this motivates the next section.
\subsection[Characterizing Redundancy: the Pattern Case]
           {Using Redundant Positions for Characterizing Redundancy: \linebreak the Pattern Case}
\label{SecUsingRedPosCharRedPatternCase}

In the following,
we provide a different sufficient criterion for redundancy
which is less demanding regarding the shape of the left hand sides,
although it requires
confluence
and
$\Seval_\cR$-definedness,
in return. 
The following definitions are helpful to determine the redundancy
of argument $i$ of $f$ when $f$ is defined by  
`matching cases' for the argument $i$ in the different rules.

\begin{definition}\label{DefUnifyUpToIthArg}
Let $\Symbols$ be a signature,
$t=f(t_1,\ldots,t_{k})$, $s=f(s_1,\ldots,s_{k})$
be terms  and $i\in\{1,\ldots,k\}$.
We say that $t$ and $s$
{\em unify up to $i$-th argument with {\em mgu} $\sigma$}
if
	\linebreak
$\langle t_1,\ldots,t_{i-1}, t_{i+1},\ldots,t_{k}\rangle$ and
$\langle s_1,\ldots,s_{i-1},s_{i+1},\ldots,s_{k}\rangle$
unify mith mgu $\sigma$.
\end{definition}

\begin{definition}[$(f,i)$-triple]
\label{Deffituple}
Let $\cR\,{=}\,(\Symbols,R)$ be a TRS,$\,f\,{\in}\:\Symbols$,$\,$and $i\,{\in}\:\{1,\ldots,ar(f)\}$.
Given two different
(possibly renamed) rules
$l\to r$, $l'\to r'$ in $\cR_f$
such that
$\Var(l)\cap\Var(l')=\emptyset$,
we say that
$\langle l\to r,l'\to r',\sigma\rangle$ is an $(f,i)$-triple of $\cR$ if
$l$ and $l'$ unify up to $i$-th argument with
{\em mgu} $\sigma$.
\end{definition}

\begin{example}\label{ExTuple}
Consider the TRS $\cR$ from Example \ref{applast}.
This program has a single $(\pr{lastnew},2)$-triple:
\[\small
\langle ~
\begin{array}[t]{@{}l@{}r@{}}
\pr{lastnew(x,nil,z)=z},~ 
 \pr{lastnew(x',y:ys,z')=lastnew(y,ys,z')},~ 
 {[\pr{x}\mapsto\pr{x'},\pr{z}\mapsto\pr{z'}]} 
  \rangle
\end{array}
\] 
\end{example}

The following definition allows us to consider rules for symbol $f$
which are ``semantically equivalent'' after replacing some variables and 
$i$-parameters in their rhs's. The basic idea is to check joinability 
of the $(f,i)$-triples of Definition \ref{Deffituple} where variables 
below the $i$-th argument of symbol $f$ in the left-hand sides of the 
rules of the triple are explicitly instantiated by a {\em dummy} 
symbol $a$ (Definition \ref{DefJoinableTriple} below). Intuitively, 
joinabilty of (all) such triples, then, amounts at proving the $i$-th 
argument of $f$ as redundant (Theorem \ref{CoroRedAndInfEvalComp} below).

In the following, we will use notation $\ol{t}$ either for a $k$-tuple of
terms $t_1,\ldots,t_k$ or for a sequence of a unique term $t,\ldots,t$;
the distinction will be clarified by the context.

\begin{definition}[Joinable $(f,i)$-triple]\label{DefJoinableTriple}
Let $\cR$ be a TRS, 
$f\in\DSymbols$, and $i\in\{1,\ldots,ar(f)\}$.
Let $a$ be an arbitrary constant.
An $(f,i)$-triple
$\langle l\to r,l'\to r',\sigma\rangle$ of $\cR$ is joinable
if  $\sigma_{\CSymbols}(\tau_{l}(r))$ and $\sigma_{\CSymbols}(\tau_{l'}(r'))$ are
joinable (i.e., they have a common reduct). Here, substitution
 $\sigma_{\CSymbols}$ is given by: 
\[
\sigma_{\CSymbols}(x)=\left \{
\begin{array}{@{}cl@{}}
\sigma(x) & \mbox{if }x\not\in\Var(l|_{i})\cup\Var(l'|_{i})\\
a & \mbox{otherwise}
\end{array}
\right .
\]
and transformation $\tau_l$ is given by
\[
\tau_l(t)=\left\{
\begin{array}{@{}cl@{}}
t 
   & \mbox{if } l|_i\in\Variables\\
t[\ol{a}]_Q
   & \mbox{if } l|_i\not\in\Variables
     \mbox{ and } Q=\{p\in Pos_{f,i}(t) \mid \Var(t|_p)\cap\Var(l|_i)\neq\emptyset\}
\end{array}
\right.
\]
\end{definition}
\noindent
Note that the constant $a$ in the previous definition can be replaced by any ground term.
In the case of many-sorted signatures, we would consider
 different constants `$a$', one for each sort.

\begin{example}\label{ExRedTuple}
Consider again the CS $\cR$ in Example \ref{applast}
and the single $(\pr{lastnew},2)$-triple
 given in Example \ref{ExTuple}.
Let us call the rhs's 
\[r_1=\pr{z} \mbox{ and }r_2=\pr{lastnew(y,ys,z')}\]
for the lh's
$l_1=\pr{lastnew(x,nil,z)}$ and $l_2=\pr{lastnew(x',y:ys,z')}$.
Let us consider that 
\pr{0} is the constant for the sort of the first argument of \pr{lastnew}
and \pr{nil} is the constant for the sort of the second argument of \pr{lastnew}.
The corresponding transformed rhs's are
\[\tau_{l_1}(r_1)=\pr{z}\mbox{ and }\tau_{l_2}(r_2)=\pr{lastnew(y,nil,z')}.\]
With $\sigma=[\pr{x}\mapsto\pr{x'},\pr{z}\mapsto\pr{z'}]$ and 
$\sigma_\CSymbols=[\pr{x}\mapsto\pr{x'},\pr{z}\mapsto\pr{z'},\pr{y}\mapsto\pr{0},\pr{ys}\mapsto\pr{nil}]$,
the corresponding instantiated rhs's
are 
\[\sigma_\CSymbols(\tau_{l_1}(r_1))=\pr{z'}\mbox{ and }\sigma_\CSymbols(\tau_{l_2}(r_2))=\pr{lastnew(0,nil,z')}.\]
We can prove 
$\sigma_\CSymbols(\tau_{l_1}(r_1))$ and $\sigma_\CSymbols(\tau_{l_2}(r_2))$
are joinable, since the variable \pr{z'} is the common reduct. 
Hence, the considered $(\pr{lastnew},2)$-triple  is
joinable.
\end{example}

Roughly speaking, the result below formalizes a method to
determine redundancy w.r.t.\ $\Seval_\cR$ which is based on finding a
common reduct of (some particular instances of) the right-hand sides
of rules.

\begin{definition}[$(f,i)$-joinable TRS]\label{joinableTRS}
Let $\cR$ be a TRS, 
$f \in \Symbols$, and $i \in \{1,\ldots,ar(f)\}$.
$\cR$ is $(f,i)$-joinable
if, for all $l \to r \in \cR_f$
and $x \in \Var(l|_i)$,
$x$ is $(f,i)$-redundant
in $r$ and all  $(f,i)$-triples of $\cR$ are joinable. 
\end{definition}

\noindent
The following result is auxiliary for Theorem \ref{CoroRedAndInfEvalComp}
and proves 
that the same constructor term is obtained by rewriting
when we replace the set of subterms
at $\Seval_\cR$-redundant and $Pos_{f,i}$ positions in a term
by an arbitrary set of terms.

\begin{proposition}\label{PropLemOrthoRedTupleGeneral}
Let
$\cR$ be a left-linear, confluent, and $\Seval_\cR$-defined CS.
Let $f \in \DSymbols$ and $i\in\{1,\ldots,ar(f)\}$.
Let $t\in\GTerms$,
$P\subseteq Pos_{f,i}(t)\cup\rpos{}{\Seval_\cR}{t}$
be a set of disjoint positions,
and $a$ be a constant.
Let $t \to^* \delta$ for some $\delta\in\GCTerms$.
If
 $\cR$ is
$(f,i)$-joinable, then 
$t[\ol{a}]_P \to^* \delta$.
\end{proposition}

Now, we provide the third effective method to detect
redundancy.

\begin{theorem}[Detecting redundancy: the Pattern Case]\label{CoroRedAndInfEvalComp}
Let $\cR$ be
a left-linear, confluent
and
$\Seval_\cR$-defined CS. Let
$f \in \DSymbols$ and 
\linebreak
$i\in\{1,\ldots,ar(f)\}$.
If
 $\cR$ is
$(f,i)$-joinable, then $i\in\rarg{\cR}{\Seval_\cR}{f}$.
\end{theorem}

\noindent
Confluence and $\Seval_\cR$-definedness are necessary, as shown in the following examples.

\begin{example}
Consider the following non-confluent CS $\cR$:
\startprog
f(0) = 0      f(s(x)) = g(f(x))      g(x) = 0      g(x) = s(0)
\stopprog
By Theorem \ref{CoroRedAndInfEvalComp}, 
we would have $1\in\rarg{\cR}{\Seval_\cR}{\pr{f}}$, since the 
$(\pr{f},1)$-triple
	\linebreak
$\tuple{\pr{f(0)=0},\ \pr{f(s(x))=g(f(x))},\ id}$
is joinable, i.e., the common reduct of 
terms \pr{0} and \pr{g(f(0))} is \pr{0}.
However, 
$\Seval_\cR(\pr{f(0)})=\{\pr{0}\}\neq\{\pr{0},\pr{s(0)}\}=\Seval_\cR(\pr{f(s(0))})$.
\end{example}

\begin{example}
Consider the following non-$\Seval_\cR$-defined CS $\cR$:
\startprog
f(0) = 0        f(s(x)) = f(x)     g(s(0)) = 0
\stopprog
By Theorem \ref{CoroRedAndInfEvalComp}, 
we would have
$1\in\rarg{\cR}{\Seval_\cR}{\pr{f}}$. 
But $\Seval_\cR(\pr{f(0)})=\{0\} \neq 
\emptyset = \Seval_\cR(\pr{f(g(0))})$.
\end{example}

Joinability is decidable for terminating, confluent
TRSs as well as for other classes of TRSs such as right-ground
TRSs \cite{Oya90} and confluent semi-constructor TRSs \cite{MOOY04}
(a semi-constructor TRS is such a TRS that every subterm of the rhs of each rewrite rule 
is ground if its root is a defined symbol).
Hence, Theorem \ref{CoroRedAndInfEvalComp} gives us
an effective method to recognize redundancy in completely defined,
confluent, and (semi-)complete TRSs, as illustrated
in the following.

\begin{example}\label{ExLastnewAndSecondSufCondRed}
Consider again the CS $\cR$ of Example \ref{applast}.
This program is confluent, terminating and completely defined
(considering sorts),
hence is $\Seval_\cR$-defined. 
By Example \ref{applast:lastnew:1},
the first argument of \pr{lastnew} is redundant w.r.t.\
$\Seval_{\cR}$, using Theorem \ref{TheorRedAndVarInRhs}.
Now, the second argument of \pr{lastnew} is redundant w.r.t.\
$\Seval_{\cR}$ using the new Theorem \ref{CoroRedAndInfEvalComp}.
As a consequence, the positions of variables \pr{x} and \pr{xs}
in the rhs of the first rule of
\pr{applast} have been proven
redundant.
Then, since
both \pr{lastnew(0,nil,z)}
and \pr{z} rewrite to \pr{z},   $\cR_{\fpr{applast}}$ is
$(\pr{applast},1)$-joinable. And again by Theorem \ref{CoroRedAndInfEvalComp},
we conclude that the first argument of \pr{applast} is also redundant.
Hence, 
$1\in\rarg{}{\Seval_\cR}{\pr{applast}}$ and
$1,2\in\rarg{}{\Seval_\cR}{\pr{lastnew}}$.
\end{example}

\begin{table}
\caption{Summary of Results} \label{tabla1}
\centering 
{\small
\begin{tabular}{clc} 
\hline\hline
Semantics & Theorem & Requirements\\ \hline
$\Sfunc$ & Th. \ref{TheoRedundancyIsAntiMonotone} (Antimonotonicity) &  --\\
$\Sfunc$ & Prop. \ref{PropThereIsNoRedArgForConstructors} (Non-redundancy) &  --\\
$\{\Sred,\Shnf,\Snf,\Seval\}$ & Th. \ref{CoroDecidRed} (Decidability) &  LL, RG\\
$\{\Snf,\Seval\}$ & Coro. \ref{CoroApproximationOfRedundancyOfEval} (Approximation $\cR_{rg}$) & LL, ND, NDT (ED, EDT) \\
$\Sfunc$ & Th. \ref{TheoRedundancyPositionIsAntiMonotone} (Antimonotonicity -- positions) &  --\\
$\Sfunc$ & Prop. \ref{PropNoRposCTerm} (Non-redundancy -- positions) &  --\\
$\Seval$ & Th. \ref{TheorRedAndVarInRhs} (The Variable Case)  & CS, LL, VR \\
$\Seval$ & Th. \ref{CoroRedAndInfEvalComp} (The Pattern Case) &  C, CS, ED, LL, JT\\ \hline
\end{tabular}}\\[.25cm]
{\footnotesize
\begin{tabular}{l@{\ :}ll@{\ :}l}
$C$ & Confluence                             & $LL$ & Left-Linearity of the TRS\\
$CS$ & Constructor System                    & $ND$ & $\Snf$-definedness\\
$ED$ & $\Seval$-definedness                  & $NDT$ & $\Snf_{\cR_{rg}}$-determinacy\\
$EDT$ & $\Seval_{\cR_{rg}}$-determinacy      & $RG$ & Right-ground TRS\\
$JT$ & Joinability of $(f,i)$-triples        & $VR$ & Variables in $l|_i$ are $(f,i)$-redundant in r\\
\hline\hline
\end{tabular}}
\end{table}

Let us conclude with a few general remarks about the complexity of our
approach, that is, the analysis time to detect redundant arguments
(the cost of performing the optimizations proposed in Section \ref{SecErasingRedArgs}  is negligible).
In Table \ref{tabla1}, we provide a summary of the main results in the paper.
Theorem \ref{TheorRedAndVarInRhs} only requires 
syntactic properties
which can be
tested in linear time on the size of the TRS (i.e., on the sum of sizes of each
rule, where the size of
a rule is the sum of sizes of the left- and right-hand sides).
The conditions $LL$, $C$ and $ED$ in the premises of Theorem \ref{CoroRedAndInfEvalComp}
are standard properties of rewrite systems 
(as remarked in Section \ref{SecRewSemantics}, a TRS $\cR$ is
$ED$ 
if $\cR$ is normalizing and completely defined,
but there is no direct way to check whether a TRS is normalizing and then termination is required)
and then assumed
to be fulfilled by the TRS $\cR$ and checked apart.
The complexity of such properties for 
decidable cases has been investigated
elsewhere (see, e.g.,
\cite{GodTiw_DecFundPropTRS_IJCAR04,KapNarRosZha_SuffCompGrouReducComplexity_AI91,Verma_AlgRedForRewProblemsII_IPL02})
and a number of tools are available for checking them in practice:
For instance,
termination tools such as \aprove\
\cite{GieThiSchFal_AProVE_RTA04}
and \cime\ \cite{ConMarMonUrb_ProvingTermOfRewWithCiME_WST03},
confluence checking tools such as \cime,
and
tools for ensuring completely-definedness such as {\sc Scc}
\cite{HendrixClavelMeseguer_RTA05_SCC}.
Thus, the only property which is strictly new in our framework is $JT$.
As we mentioned above, joinability is decidable for several classes of TRSs
\cite{GodTiw_DecFundPropTRS_IJCAR04,Verma_AlgRedForRewProblemsII_IPL02,MOOY04}.
Actually, there are (cubic) polynomial time algorithms for joinability of ground
systems
\cite[Theorem 12]{Verma_AlgRedForRewProblemsII_IPL02}
and 
a slightly more general class of TRSs is
considered in
\cite{GodTiw_DecFundPropTRS_IJCAR04}, namely right-(ground or variable)
rewrite systems. 
In our implementation however,
confluence and termination of the TRS are assumed for the application of Theorem \ref{CoroRedAndInfEvalComp}
(see above) and then
joinability of terms $t$ and $s$ is decidable
by just checking whether the normal forms of $t$ and $s$ are equal.

\section{Erasing Redundant Arguments}\label{SecErasingRedArgs}

The presence of redundant arguments
within input expressions wastes memory space and
can lead to time consuming explorations and transformations (by
replacement) of their structure.
Then, since redundant arguments are not necessary to determine 
the result of a function call, it is worth to develop methods and 
techniques to avoid such unpleasant effects.

As remarked in the introduction, inefficiencies caused by the redundancy 
of arguments cannot (in general) be avoided by using rewriting strategies.
In this section we formalize a procedure for
{\em removing} redundant arguments from a TRS. The basic idea is simple: if an
argument of $f$ is redundant, it does not contribute to obtaining the
value of  any call to $f$ and can be dropped from program $\cR$.
Hence, we remove redundant formal parameters and corresponding
actual parameters for each function symbol and function call in $\cR$.
We begin with the notion of syntactic erasure which is intended to
pick up redundant arguments of function symbols.

\begin{definition}[Syntactic erasure]%
A {\em syntactic erasure} is a mapping
$\rho:\Symbols\to \pwset(\nat)$ such that for all
$f\in\Symbols,~\rho(f)\subseteq\{1,\ldots,ar(f)\}$. We say that a
syntactic erasure $\rho$ is {\em sound} for a semantics
$\Sfunc$ if, for all $f\in\Symbols$, $\rho(f)\subseteq\rarggen$.
\end{definition}

\begin{example}\label{ExSoundSyntacticErasure} 
Given  the signature 
$\Symbols=\{\pr{0},\pr{nil},\pr{s},\pr{:},\pr{applast},\pr{lastnew}\}$
of the TRS $\cR$ in Example \ref{applast}, 
with $ar(\pr{0})\,{=}\,ar(\pr{nil})\,{=}\,0$, $ar(\pr{s})\,{=}\,1$,
$ar(\pr{:})\,{=}\,ar(\pr{applast})=2$,
and $ar(\pr{lastnew})=3$, and according to Example 
\ref{ExLastnewAndSecondSufCondRed},
the following mapping $\rho$ is
 a {\em sound syntactic erasure} for the semantics $\Seval_\cR$:
$
\rho(\pr{0})=\rho(\pr{nil})=\rho(\pr{s})=\rho(\pr{:})=\emptyset,\ 
\rho(\pr{applast})=\{1\}, \mbox{ and } 
\rho(\pr{lastnew})=\{1,2\}
$.
\end{example}

Since we are interested in {\em removing} redundant arguments from
function symbols, we transform the functions by reducing
their arity  according
to the information provided by the redundancy analysis, thus building a new,
 {\em erased}
signature.

\begin{definition}[Erasure of a signature]
Given a
signature $\Symbols$ and a syntactic erasure
$\rho:\Symbols\to \pwset(\nat)$, the erasure of
$\Symbols$ is the signature  $\Symbols_\rho$  whose symbols
$f_\rho\in\Symbols_\rho$ are one to one with symbols $f\in\Symbols$ and
whose arities are related by $ar(f_\rho)=ar(f)-|\rho(f)|$.
\end{definition}

\begin{example}\label{ExSignaErased} 
The erasure of the signature in Example \ref{ExSoundSyntacticErasure}
is
$\Symbols_{{\rho}}=\{\pr{0},\pr{nil},\pr{s},\pr{:},
\pr{applast},
\pr{lastnew}\}$, with $ar(\pr{0})=ar(\pr{nil})$
$=0$, $ar(\pr{s})=ar(\pr{applast})=
ar(\pr{lastnew})=1$, and
$ar(\pr{:})=2$.
Note that, by abuse, we use the same symbols for the functions of the erased
signature.
\end{example}

\noindent
Now we extend the procedure to terms in the obvious way.

\begin{definition}[Erasure of a term]\label{DefErasureOfaTerm}
Given a
syn\-tac\-tic erasure $\rho:\Symbols\to \pwset(\nat)$,
the function $\tau_\rho:\Terms\to\TermsRho$ on terms is:
$\tau_\rho(x)=x$ if $x\in\Variables$ and
$\tau_\rho(f(t_1,\ldots,t_n))=
f_\rho(\tau_\rho(t_{i_1}),\ldots,\tau_\rho(t_{i_k}))$ where
$\{1,\ldots,n\}-\rho(f)=\{i_1,\ldots,i_k\}$ and $i_m<i_{m+1}$ for $1\leq
m< k$.
\end{definition}

\noindent
The erasure procedure is extended to TRSs: we erase the
lhs's and rhs's of each rule according to $\tau_\rho$. In order to
avoid extra variables in rhs's of rules (that arise from
the elimination of redundant arguments of symbols in the corresponding lhs),
we replace them by an arbitrary constant of $\Symbols$ (which
automatically belongs to $\Symbols_\rho$).

\begin{definition}[Erasure of a TRS]
\label{DefPreErasureOfaTRS}
Let $\cR$ be a TRS, $a$ a constant, and $\rho$ be a syntactic erasure for $\Symbols$. 
The erasure
$\Rerased$ of $\cR$ is $\Rerased=(\Symbols_\rho,
\{\tau_\rho(l)\to \sigma_l(\tau_\rho(r))~|~l\to r\in R\})$
where the substitution $\sigma_l$ for a $lhs$ $l$ is given by
$\sigma_l(x)=a$ for all
$x\in\Var(l)-\Var(\tau_\rho(l))$ and $\sigma_l(y)=y$ whenever
$y\in\Var(\tau_\rho(l))$.
\end{definition}
\noindent
Note that the constant $a$ in the previous definition can be replaced by any ground term.
In a many-sorted signature, we will have different constants `$a$', each one of an appropriate sort.

\begin{example}\label{ExSyntErasure}
Let $\cR$ be the TRS of Example \ref{applast} 
and
$\rho$ be the sound syntactic erasure of Example
\ref{ExSoundSyntacticErasure}.
The erasure $\cR_\rho$ of $\cR$ consists of the erased signature of
Example \ref{ExSignaErased} together with the following rules:
\startprog
applast(z) = z           lastnew(z) = z\nopagebreak
applast(z) = lastnew(z)  lastnew(z) = lastnew(z)
\stopprog
Below, we introduce a further improvement aimed at obtaining
the final, ``optimal'' program.
\end{example}
The mapping $\tau_\rho$ induces an equivalence $\equiv_{\tau_\rho}$
on terms given by: $t\equiv_{\tau_\rho}s$ iff
$\tau_\rho(t)=\tau_\rho(s)$. 
We have the following property of
sound erasures of terms.

\begin{proposition}\label{PropEquivErasurePreservesSemantics}
If the syntactic erasure $\rho:\Symbols\to \pwset(\nat)$ is sound with
respect to the semantics $\Sfunc$, then for all $t,s\in\GTerms$,
$t\equiv_{\tau_\rho}s$ implies that $\Sfunc(t)=\Sfunc(s)$.
\end{proposition}

The following theorem establishes the correctness of the erasure procedure
for a rewriting semantics $\Sfunc$.

\begin{theorem}[Correctness]\label{TheoCorrPreErasure}
Let $\cR$ be a left-linear TRS, $\Sfunc$ be a rewriting
semantics for $\cR$, $\rho$ be a sound
syntactic erasure for $\Sfunc$, and $t\in\GTerms$. If
$\delta\in\Sfunc(t)$, then
$\tau_\rho(t)\to^*_{\cR_\rho}\tau_\rho(\delta)$.
\end{theorem}

The following theorem establishes the completeness of the erasure procedure
for a rewriting semantics $\Sfunc$.

\begin{theorem}[Completeness]\label{TheoCompPreErasureEval}
Let $\cR$ be a left-linear TRS,
$\Sfunc$ be a rewriting semantics for $\cR$
such that $\Sfunc \preceq \Sred_{\cR}$,
$\rho$ be a sound syntactic erasure for $\Sfunc$,
and $t,\delta\in\cT(\Symbols_\rho)$.
If $t\to^*_{\cR_\rho}\delta$, then
$\forall t',\delta'\in\GTerms$ such that $\tau_\rho(t')=t$ and
$\tau_\rho(\delta')=\delta$,
$\Sfunc(\delta')\subseteq\Sfunc(t')$.
\end{theorem}

The following theorem establishes the correctness and completeness of
the erasure procedure for the semantics $\Seval_\cR$.

\begin{theorem}[Correctness and Completeness]
\label{TheoCorrPreErasureEval}
Let $\cR$ be a left-linear TRS,
 $\rho$ be a sound syntactic erasure for $\Seval_\cR$,
 $t\in\GTerms$, and $\delta\in\GCTerms$. Then,
$\tau_\rho(t)\to^*_{\cR_\rho}\delta$ iff $\delta\in\Seval_\cR(t)$.
\end{theorem}

In the following, we are able to 
ascertain the conditions for the preservation of
some computational properties of TRSs after erasure.

\begin{theorem}[Preservation of Confluence]\label{TheoConfluentRho}
Let $\cR$ be a left-linear TRS.
Let $\rho$ be a sound syntactic erasure for $\Seval_\cR$.
If $\cR$ is
$\Seval_\cR$-defined
and confluent, then the erasure
$\cR_\rho$ of $\cR$ is confluent.
\end{theorem}

\begin{theorem}[Preservation of Normalization]\label{TheoWeakTermRho}
Let $\cR$ be a left-linear and completely defined TRS, and
$\rho$ be a sound syntactic erasure for $\Seval_\cR$.
If $\cR$ is normalizing, then the erasure $\cR_\rho$ of $\cR$ is normalizing.
\end{theorem}

In the theorem above, we cannot strengthen normalization to
termination. A simple counterexample showing that termination
may get lost is the following; note that the opposite is also possible, 
i.e., a non-terminating TRS can be made terminating after the erasure.

\begin{example}
Consider the left-linear, (confluent, completely defined, and) terminating TRS $\cR$
\startprog
h(a,y) = a    h(c(x),y) = h(x,c(y))
\stopprog
The first argument of \pr{h} is redundant
w.r.t.\ $\Seval_\cR$.
However, after erasing the argument, we get the 
TRS
\startprog
h(y) = a      h(y) = h(c(y))
\stopprog
which is not terminating.
\end{example}

\noindent
In the example above, note that the resulting TRS is not orthogonal,
whereas the original program is. Hence, this example also shows that
orthogonality is not preserved under  erasure.

After the erasure, 
a post-processing transformation able to remove redundant rules 
(w.r.t. an appropriate notion of rule redundancy) 
might be useful to restore termination or orthogonality in some cases, as the example above. Although this point is outside the scope of this paper,
in the following we provide a program transformation 
that can improve the optimization achieved by the erasure.

\begin{definition}[Reduced erasure of a TRS]
\label{DefErasureOfaTRS}
Let
$\cR$ be a
TRS and
$\rho$ be a syntactic erasure for $\Symbols$. The reduced erasure
$\cR'_\rho$ of $\cR$ is obtained from the erasure $\cR_\rho$ of
$\cR$ by a {\em compression transformation} defined as
removing any trivial rule $t\to t$ of  $\cR_\rho$  and then normalizing
the rhs's of the rules w.r.t.\ the non-trivial rules of $\cR_\rho$.
\end{definition}

Reduced erasures are well-defined whenever $\cR_\rho$ is confluent and
normalizing since, for such systems, every term has a unique normal form.

\begin{example}\label{ExReducedErasureTRSApplast}
Let $\cR_\rho$ be the erasure of Example \ref{ExSyntErasure}. 
The reduced erasure consists of the 
rules
$\{\pr{applast(z) = z},\ \pr{lastnew(z) = z}\}$.
\end{example}

Since right-normalization preserves confluence, termination and
the equational theory (as well as confluence, normalization and
the equational theory, in almost orthogonal and normalizing TRSs)
\cite{Gramlich_OnIntSemiCompTRSs_TCS01},
and the removal of trivial rules does not change the evaluation semantics
of the TRS $\cR$ either, we have the following.

\begin{corollary}
\label{TheoCompCorrErasureEval}
Let $\cR$ be a left-linear TRS, $\rho$ be a sound syntactic erasure for
$\Seval_\cR$, $t\in\GTerms$, and
$\delta\in\GCTerms$. If (the TRS which results from
removing trivial rules from) $\cR_\rho$ is confluent and terminating
(alternatively, if it is almost
orthogonal and normalizing), then,
$\tau_\rho(t)\to^*_{\cR'_\rho}\delta$ if and only
if $\delta\in\Seval_\cR(t)$, where ${\cR'_\rho}$ is the reduced erasure of
{\cR}.
\end{corollary}
Erasures and reduced erasures of a TRS preserve left-linearity.
For a TRS $\cR$ satisfying the conditions in Corollary
\ref{TheoCompCorrErasureEval}, by using
\cite{Gramlich_OnIntSemiCompTRSs_TCS01}, it is immediate that
the reduced erasure $\cR'_\rho$
is confluent and normalizing. Also, $\cR'_\rho$
is completely defined if $\cR$  is.

Hence, let us note that these results   allow us to perform the
`optimal' optimization of program \pr{applast} in Example \ref{applast}
while guaranteeing that the intended 
(evaluation or normalization)
semantics is preserved.

\section{Experiments}\label{SecExperiments}

The practicality of our ideas is witnessed by the implementation
of a prototype system which delivers encouraging good results
for
the techniques deployed in 
Section \ref{SecCharRed} (Theorems \ref{TheorRedAndVarInRhs} and \ref{CoroRedAndInfEvalComp})
and the erasure procedure of Section 
\ref{SecErasingRedArgs}.
The prototype
has been implemented in \Pakcs\ \cite{PAKCS03}, 
the current distribution\footnote{See \url{http://www.informatik.uni-kiel.de/~pakcs}} of the 
multi-paradigm declarative language \Curry\ \cite{CurryTR},
and is publicly available at 
\url{http://www.dsic.upv.es/users/elp/redargs}.

We have used the prototype to perform some
preliminary experiments 
which show that
our methodology
does detect and remove redundant arguments
of  some common transformation benchmarks,
such as \pr{bogus}, \pr{lastappend}, \pr{allzeros}, 
	\linebreak
\pr{doubleflip}, etc.;
see \cite{Leu98} and references therein.
Tables \ref{FigTimesRedArgsCurry} and \ref{FigTimesRedArgsMaude}
summarize the experiments. 
Benchmarks code as well as the programs obtained by the erasure procedure are 
included in 
\ref{AppCharRedPos}. 

Table \ref{FigTimesRedArgsCurry} shows the execution runtimes of the original and transformed programs 
in \Pakcs, as well as the arguments in the whole program which are signaled as redundant for 
each benchmark using the notation: \#signaled/\#total.
Runtimes have been measured in an ``AMD Athlon XP'' class machine running {\sf Fedora Core} $3.0$
and using version 1.6.0 of the \Pakcs\ compiler under \sicstusProlog\ 3.8.6.
Natural numbers are given by numbers $0$, $1$, $2$, etc in the tables,
instead of the notation \pr{Z}/\pr{S} x used in the bechmarks code. 
For benchmarking purposes, goals make use of the auxiliary factorial function, defined in a usual way.
The number of elements of a list (when used) is indicated by a subindex. 
Note that the analysis time for each example is negligible.

Important optimizations are obtained for most examples.
In the case of program \pr{bogus}, no appreciable optimization is achieved by removing 
redundant arguments, since \Curry\ is a lazy language and the redundant argument in \pr{bogus} is a useless variable.
In order to dissociate the possible dependency of the achieved optimization 
w.r.t.\ the lazy evaluation of the language, Table 
\ref{FigTimesRedArgsMaude} shows the execution runtimes 
of the benchmarks in the \Maude\ interpreter\footnote{See \url{http://maude.cs.uiuc.edu}}
(version 2.1.1),
which uses an innermost rewriting strategy.

Note that, in this case, significant optimizations are also measured for programs 
\pr{bogus} and  \pr{applast}.
The \pr{plus\_minus} example
runs in nearly half the original execution time in both, lazy and eager systems,
which seems consistent with the fact that one of the two arguments have been removed.

\begin{table}
\caption{Execution of the original and transformed programs in \Curry}
\label{FigTimesRedArgsCurry}
\begin{tabular}{@{}ll@{}c@{}c@{}c@{}}
\hline\hline
Name
&
Call in original/erased program
&
Time (ms)
&
Gain
&
${\it rarg}_{\Seval}$
\\ \hline
bogus
&
\pr{loop (fact 8) (fact 9) (fact 8)}
&
150
&
\\
&
\pr{loop' (fact 8) (fact 8)}
&
150
&
0\%
&
1/1
\\ \noalign{\vspace{.15cm}}
applast
&
\pr{applast [(fact 8)]$_{10000}$ (fact 8)}
&
168
&
\\
&
\pr{applast' (fact 8)}
&
153
&
9\%
&
3/3
\\ \noalign{\vspace{.15cm}}
plus\_minus
&
\pr{minus\_pe (fact 8) (fact 8)}
&
220
&
\\
&
\pr{minus\_pe' (fact 8)}
&
155
&
30\%
&
1/1
\\ \noalign{\vspace{.15cm}}
plus\_leq
&
\pr{leq\_pe (fact 8) (fact 8)}
&
79
&
\\
&
\pr{leq\_pe'}
&
$\sim$0
&
100\%
&
1/1
\\ \noalign{\vspace{.15cm}}
double\_even
&
\pr{even\_pe (fact 8)}
&
77
&
\\
&
\pr{even\_pe'}
&
$\sim$0
&
100\%
&
1/1
\\ \noalign{\vspace{.15cm}}
sum\_allzeros
&
\pr{sum\_pe [(fact 8)]$_{10000}$}
&
23
&
\\
&
\pr{sum\_pe'}
&
$\sim$0
&
100\%
&
1/1
\\ \noalign{\vspace{.15cm}}
Mutual recursion 1
&
\pr{f (fact 8) (fact 8)}
&
123
&
\\
&
\pr{f'}
&
$\sim$0
&
100\%
&
1/1
\\ \noalign{\vspace{.15cm}}
Mutual recursion 2
&
\pr{f (fact 8)}
&
132
&
\\
&
\pr{f'}
&
$\sim$0
&
100\%
&
1/1
\\
\hline\hline
\end{tabular}
\end{table}

\begin{table}
\caption{Execution of the original and transformed programs in \Maude}
\label{FigTimesRedArgsMaude}
\begin{tabular}{@{}llc@{}c@{}c@{}}
\hline\hline
Name
&
Call in original/erased program
&
Time (ms)
&
Gain
&
${\it rarg}_{\Seval}$
\\ \hline
bogus
&
\pr{loop (fact 8) (fact 9) (fact 8)}
&
651
&
\\
&
\pr{loop' (fact 8) (fact 8)}
&
47
&
93\%
&
1/1
\\ \noalign{\vspace{.15cm}}
applast
&
\pr{applast [(fact 8)]$_{10000}$ (fact 8)}
&
102
&
\\
&
\pr{applast' (fact 8)}
&
54
&
47\%
&
3/3
\\ \noalign{\vspace{.15cm}}
plus\_minus
&
\pr{minus\_pe (fact 8) (fact 8)}
&
62
&
\\
&
\pr{minus\_pe' (fact 8)}
&
30
&
51\%
&
1/1
\\ \noalign{\vspace{.15cm}}
plus\_leq
&
\pr{leq\_pe (fact 8) (fact 8)}
&
33
&
\\
&
\pr{leq\_pe'}
&
$\sim$0
&
100\%
&
1/1
\\ \noalign{\vspace{.15cm}}
double\_even
&
\pr{even\_pe (fact 8)}
&
32
&
\\
&
\pr{even\_pe'}
&
$\sim$0
&
100\%
&
1/1
\\ \noalign{\vspace{.15cm}}
sum\_allzeros
&
\pr{sum\_pe [(fact 8)]$_{10000}$}
&
40
&
\\
&
\pr{sum\_pe'}
&
$\sim$0
&
100\%
&
1/1
\\ \noalign{\vspace{.15cm}}
Mutual recursion 1
&
\pr{f (fact 8) (fact 8)}
&
73
&
\\
&
\pr{f'}
&
$\sim$0
&
100\%
&
1/1
\\ \noalign{\vspace{.15cm}}
Mutual recursion 2
&
\pr{f (fact 8)}
&
61
&
\\
&
\pr{f'}
&
$\sim$0
&
100\%
&
1/1
\\
\hline\hline
\end{tabular}
\end{table} 

\section{Related Work}\label{SecRelatedWork}

Some notions have appeared in the literature of what it means for a
term in a TRS $\cR$ to be ``computationally irrelevant''.
As we are going to see,
our analysis is different from all the related methods in many respects
and, in general, incomparable to them.

Contrarily to our notion of
redundancy, the meaninglessness of \cite{Kup94,KOV96} is a property of
the terms themselves (they may have meaning in $\cR$ or may not),
whereas our notion refers to arguments (positions) of function symbols.
In \cite[Section 7.1]{Kup94}, a term $t$ is called
{\em meaningless\/} if,
for each context $C[\;]$ s.t. $C[t]$ has a normal form, we have that
$C[t']$ has the same normal form for all terms $t'$.
This can be seen as a kind of superfluity (w.r.t.\ normal forms)
of a fixed expression
in any context, whereas our notion of
redundancy refers to the possibility of getting rid of some arguments
of a given function symbol with regard to some observed semantics.
The meaninglessness of \cite{Kup94} is not helpful
for the purposes of optimizing
programs by removing useless arguments of function symbols which we
pursue.
On the other hand, terms with a normal form are proven
meaningful (i.e., not meaningless) in \cite{Kup94,KOV96},
whereas we might have redundant 
actual parameters 
which are normal forms.

Among the vast literature on analysis (and removal) of unnecessary data
structures,
the analyses of {\em unneededness} (or {\em absence})  
of functional programming
\cite{CousotCousot94-1,hughes:backwards},
and the {\em filtering} of useless arguments and unnecessary variables
of logic programming
\cite{LeuschelSorensen:RAF,PP94} are the closest to our work.
In \cite{hughes:backwards}, a notion of {\em needed/unneeded}
parameter for list-manipulation programs is introduced which
is closely related to the redundancy of ours in that it is
capable of identifying whether the value of a subterm is ignored.
The method is formulated in terms of a fixed,
finite set of projection functions which introduces some limitations
on the class of neededness patterns that can be identified.
Since our method
gives the information that a parameter is definitely not necessary,  our
redundancy notion implies Hughes's unneededness, but not vice versa.
For instance, constructor symbols cannot have redundant arguments in
our framework
(Proposition \ref{PropThereIsNoRedArgForConstructors}),
whereas Hughes' notion
of unneededness can be applied to the elements of a list,
as shown in the following example.

\begin{example}\label{ExaLength}
Consider the following TRS defining the length function for lists.%
\startprog
length(nil) = 0                length(x:xs) = s(length(xs))    
\stopprog
Hughes' analysis is able to determine
that, in the \pr{length} function, 
the spine of
the argument list is needed but the elements of the list are not
needed; this is used to perform some optimizations for the compiler.
However, this information cannot  be used for the purposes of our
 work, that is,
to remove  these elements when the entire list cannot be eliminated.
\end{example}

On the other hand, Hughes's notion of {\em neededness/unneededness}
should not be
confused with the standard notion of needed (positions of) redexes
of \cite{HL91}: Example
\ref{ExNeededAndRedundancy} shows that
Huet and Levy's neededness does not imply the non-redundancy of
the corresponding  argument or position (nor vice versa).

The notion of redundancy of an argument in a term rewriting system
can be seen as a kind of {\em comportment property} as defined
in \cite{CousotCousot94-1}. Cousot's comportment analysis generalizes
not only the unneededness analyses
but also strictness, termination and other standard analyses of
functional programming.
In \cite{CousotCousot94-1},
comportment is mainly investigated within a denotational framework,
whereas  our approximation is independent from the semantic formalism.

Proietti and Pettorossi's
{\em elimination procedure\/} for the
removal of unnecessary variables is a powerful unfold/fold-based
transformation procedure for logic programs; therefore, it does not compare
directly with our method, which would be seen
as a post-processing phase for program transformers optimization.
Regarding the kind of {\em unnecessary variables\/} that the elimination procedure
can remove, only variables that occur more than once
in the body of the program rule and which do not occur in the head
of the rule
can be dropped. This is not to say that the
transformation is powerless; on the contrary, the
effect can be very striking as these kinds of variables
often determine multiple traversals of intermediate data structures
which are then removed from the program.
Our procedure for removing redundant arguments is also related to the Leuschel
and S{\o}rensen RAF and FAR algorithms \cite{LeuschelSorensen:RAF}, which
apply to removing unnecessary arguments
in the context of (conjunctive) partial evaluation
of logic programs.
 However, a comparison is not easy either
as we have not yet considered the semantics of computed answers
for our programs in detail.

People in the functional programming community have also studied
the problem of useless variable elimination (UVE).
Apparently, they were unaware of the works of the logic programming
community,
and they started studying the topic from scratch, mainly
following a flow-based approach \cite{WS99} or
a type-based approach \cite{BCDG00,Kobayashi00};
see \cite{BCDG00} for a discussion
of this line of research.
All these works address the problem of safe elimination of dead {\em
variables}
but heavily handle data structures.
A notable exception is \cite{LS02},
where Liu and Stoller discuss how to safely eliminate dead code
in the presence of recursive data structures by applying
a methodology based on regular tree grammars.
Unfortunately, the method in \cite{LS02} does not apply to
achieve the optimization pursued  in our running example \pr{applast}.

Obviously, there 
exist examples (inspired) in the previously discussed works
which cannot be directly handled with our results.

\begin{example}
Consider the TRS of Example \ref{ExaLength} together with the following function
symbol \pr{f}:%
\startprog
f(x) = length(x:nil)
\stopprog
Our methods do not capture the redundancy of the argument of \pr{f}.
In \cite{LS02} it is shown that, in order to evaluate \pr{length(xs)}, we
do not need to evaluate the elements of the argument list \pr{xs};
as Hughes's unneededness. In
Liu et~al.'s methodology,
this means that we could replace the rule for \pr{f} above by
the rule 
\pr{f(\_) = length(\_:nil)}
where \pr{\_} is a new (dummy) constant.
Nevertheless,
the new TRS 
can be used now 
to recognize the first argument of \pr{f} as redundant. That is, we are
allowed to use the following rule 
\pr{f = length(\_:nil)}
which completely avoids wasteful computations on redundant arguments.
Hence, the different methods are complementary
and an enhanced
test might be developed by properly combine them.
\end{example}

\section{Functional Logic Programming: Narrowing}\label{SecNarrowing}

Programs written in  muti-paradigm functional-logic
languages such as Curry (see e.g.\ those in \ref{AppCharRedPos}) are usually not
different from (equivalent) programs written in the (pure) functional
language \Haskell. The difference only shows up during the evaluation.
In Curry, one can evaluate expressions containing logical variables
(that are evaluated non-deterministically to deliver {\em computed answers} as in Prolog) while
in \Haskell\ only completely ground expressions can be (deterministically) evaluated
to compute  its {\em value}. In fact, Term Rewriting Systems are also 
used as abstract models of programs written in such languages, 
although narrowing, rather than rewriting, is usually the underlying 
computational mechanism \cite{Han94JLP}.

Before the conclusions, let us discuss how the notions and techniques 
presented so far could be adapted to cope with more sophisticated,
multi-paradigm  functional-logic languages.
The most popular operational principle to deal with logical variables within 
function calls is known as \emph{narrowing}, as used in functional logic programming 
(see \cite{Han94JLP} for a survey). 
Narrowing is an  unification-based, parameter-passing mechanism which
extends functional evaluation through goal solving capabilities as in
logic programming.
A {\em narrowing step\/} instantiates variables
of  an expression
 and then applies a reduction step to a redex of the instantiated expression.
 The instantiation of variables is usually computed by unifying
 a subterm of the entire expression with the left-hand side of some
 program equation.
Narrowing provides completeness in the sense of logic 
programming, i.e., computation of answers, as well as 
functional programming, i.e., computation of normal forms.
Formally, a  term $s$ narrows to $t$ in $\cR$,
denoted by $s \leadsto_{\sigma} t$,
iff there exists a non-variable position
$p$ of $s$, a (standardized apart) rule $l \rightarrow r \in \cR$,
and a substitution $\sigma$ such that
$s|_{p}$ and $l$ unify with mgu $\sigma$ and
$t = \sigma(s[r]_{p})$.

Narrowing can be considered as a mapping (or semantics)
$\Sfunc:\Terms\to\pwset(\Subst\times\Terms)$
that associates a set of pairs 
$\langle$substitution,term$\rangle$
to an input term \cite{HanLuc_EvSemNarrBasFLL_JFLP01}. 
The following is a typical evaluation semantics based on narrowing
$$\SevalN(t)=\{\tuple{\sigma,s} \mid t\leadsto^*_{\sigma} s \wedge s\in\CTerms\}$$
The substitutions computed by narrowing 
are usually restricted to the variables 
of the input term.
Within this semantic framework,
the idea of redundancy for term rewriting as proposed in Definition \ref{context}
cannot be na\"{\i}vely lifted to redundancy for narrowing (considering arbitrary input terms), 
as revealed by the following example.

\begin{example}\label{ExNarrowing}
Consider the TRS of Example \ref{applast}.
The first argument of symbol $\pr{lastnew}$ is redundant w.r.t. $\SevalN$,
i.e., for all contexts ${C[\;]}$ and for all $t,s\in \Terms$ such that $root(t)=f$,
 $\SevalN(C[t])=\SevalN(C[t[s]_i])$.
For instance, with the input term $t=\pr{lastnew(x,0:nil,s(0))}$,
we have $\SevalN(t[s]_1)=\{\tuple{id,\pr{s(0)}}\}$ for all $s\in\Terms$.
This is because, in every lhs, the first argument is a variable that is never inspected in the 
corresponding rhs.
However, the second argument of symbol $\pr{lastnew}$ is not 
redundant w.r.t. $\SevalN$.
Consider the goal $t'=\pr{lastnew(x,y,s(0))}$,
then 
$\SevalN(t')=\{\tuple{[\pr{y}\mapsto\pr{nil}],\pr{s(0)}},
              \tuple{[\pr{y}\mapsto\pr{w:nil}],\pr{s(0)}},\ldots\}
             \neq\{\tuple{id,\pr{s(0)}}\}=\SevalN(t'[\pr{nil}]_2)$.
The reason is that there exist many narrowing derivations for $t'$:
\[\small
\pr{lastnew(x,y,s(0))} \leadsto_{\{\fpr{y}\mapsto \fpr{nil}\}} \pr{s(0)},\ 
\pr{lastnew(x,y,s(0))} \leadsto^{\ast}_{\{\fpr{y}\mapsto \fpr{w:nil}\}} \pr{s(0)},\ 
\ldots
\]
but only this one for $t'[\pr{nil}]_2$:
$\pr{lastnew(x,nil,s(0))} \leadsto_{id} \pr{s(0)}$.
\end{example}

Thus, the general problem of analyzing redundancy w.r.t. the observable of computed 
answers  is a challenging line of research that we pursue as future work 
(hence outside the scope of this paper).
Nevertheless, we can still 
outline
different possibilities for analyzing redundancy of arguments w.r.t. narrowing 
in some particular cases by applying the results in this paper.

\paragraph{Restriction to the variable case}

We have seen in Example \ref{ExNarrowing} that
the na\"{\i}ve notion of redundancy for narrowing
is still fruitful when we consider
the case of an argument in lhs's that always corresponds to a variable
that is never inspected during the computation, i.e. the Variable Case 
in Section \ref{SecUsingRedPosCharRedVarCase}. For instance, 
using Theorem \ref{TheorRedAndVarInRhs}, we can identify
that the first argument of symbol $\pr{lastnew}$ in Example \ref{applast}
is redundant for narrowing and that also
the second argument of symbol \pr{loop} in Example \ref{Exabogus}
is redundant for narrowing. 

\paragraph{Input terms with mode information}

Since the narrowing space is bigger than the rewriting space,
the functional logic community 
(as well as the program transformation and partial evaluation community)
usually restrict their interest to preserve
the narrowing semantics $\SevalN$ for a fixed set of goals, 
similarly to the argument filtering technique of 
\cite{LeuschelSorensen:RAF}
for logic programming.

\begin{example}
Consider again the TRS of Example \ref{applast}.
Let us assume that we are only interested 
in the evaluation semantics of input terms that fit the shape 
\pr{lastnew($G$,$G$,$NG$)}, where $G$ denotes a ground term
and $NG$ an arbitrary term.
This is known as \emph{mode information} in logic programming and implies 
that the first and second arguments of symbol \pr{lastnew} are understood only as input data
whereas the third argument is understood as input and output data.
Then the techniques presented
in Section \ref{SecCharRed} can be applied to the arguments that are labeled with $G$.
For instance, the first and second arguments of \pr{lastnew} will be detected as redundant
for the mode \pr{lastnew($G$,$G$,$NG$)}.
\end{example}

As mentioned before, more research is needed in order to come upon 
a generally correct notion of redundancy w.r.t. narrowing, which leads to effective
detection algorithms that pay off in practice.
We believe that our results in this paper can be valuable for
these studies. 

\section{Conclusion}\label{SecConclusions}
This work 
provides the first results concerning
the detection and removal of useless arguments
in program functions.
We developed our results in a stepwise manner.
We have given a semantic definition of redundancy
which takes the semantics $\Sfunc$ as a parameter. 
We have considered different (reduction) semantics, 
including the standard normalization semantics
(typical of pure rewriting)
and the evaluation semantics (closer to functional programming).
We have provided some decidability results about redundancy
of an argument 
and a first effective method for detecting redundancies, which is based on
approximation techniques.
We have also provided two more practical methods
to recognize redundancy which allows us to simplify
the general redundancy problem  to the analysis of
the rhs's of the program rules.
All the three methods to detect redundancies are different and useful.
Moreover, we think that all  results in  this paper are of independent
interest and can be used for other applications in the fields of rule-based
and multi-paradigm declarative programming.

Actually, inefficiencies caused by the redundancy of arguments cannot
be avoided by using standard reduction strategies. Therefore, we have
developed a transformation for eliminating dead code which appears in
the form of useless function calls and we have proven that the
transformation preserves the semantics (and some operational properties)
of the original program under ascertained conditions.
The optimized program that we produce cannot be created as the result
of applying standard transformations of functional
programming to the original program, such as partial evaluation, supercompilation, and
deforestation, see e.g.,\ \cite{PP96}.

Furthermore, 
a prototype implementation of the (more practical) methods 
to detect redundancy together with the erasure procedure has been provided.
The preliminary experiments performed with the prototype
indicate that our 
approach is both practical and useful.
We  believe that the
semantic grounds for redundancy analyses and elimination
laid in this work 
may foster further
insights and developments in the
program optimization 
community and neighbouring fields.

Finally, apart from these comments,
the problem of identifying redundant arguments of function symbols
has been reduced to proving the validity of a particular class of
inductive theorems in the equational theory of confluent, 
$\Seval_\cR$-defined
TRSs.
We refer to \cite{AEEL02} for details, where a comparison with approximation
methods based on abstract interpretation can also be found.

\paragraph{Acknowledgements}

We thank the anonymous referees for the useful
remarks and suggestions which helped to improve the paper.

This work has been partially supported by
the EU (FEDER) and the Spanish MEC under grant TIN 2004-7943-C04-02, the
Generalitat Valenciana under grant GV03/25,
and the ICT for EU-India Cross-Cultural Dissemination 
ALA/\linebreak[0]95/\linebreak[0]23/\linebreak[0]2003/\linebreak[0]077-\linebreak[0]054 project.

\clearpage
\appendix

\section{Proofs}\label{AppProofs}

\subsection*{Proofs of Section \ref{SecRedArgInRewriting}}

\begin{ftheorem}{\ref{TheoRedundancyIsAntiMonotone}}
Let $\Sfunc,\Sfunc'$ be term semantics
for a signature $\Symbols$. If $\Sfunc\preceq\Sfunc'$, then, for all
$f\in\Symbols$, $\rarg{\Symbols}{\Sfunc'}{f}\subseteq\rarg{\Symbols}{\Sfunc}{f}$.
\end{ftheorem}

\begin{proof}
By contradiction.
Given $f\in\Symbols$ and $i\in\rarg{\cR}{\Sfunc'}{f}$, by Definition \ref{context} 
we have that, for all contexts
$C[\;]$ and for all $t,s\in \GTerms$ such that $root(t)=f$,
$\Sfunc'(C[t])=\Sfunc'(C[t[s]_i])$. Now, since $\Sfunc\preceq\Sfunc'$,
there exists $T\subseteq\GTerms$ such that
$\Sfunc(C[t])=\Sfunc'(C[t])\cap T=\Sfunc'(C[t[s]_i])\cap
T=\Sfunc(C[t[s]_i])$.
Hence, $i\in\rarg{\cR}{\Sfunc}{f}$.%
\end{proof}

\begin{fproposition}{\ref{PropThereIsNoRedArgForConstructors}}
Let $\cR$ be a TRS such that $|\GCTerms|>1$, and
consider a rewriting semantics $\Sfunc$  such that $\Seval_{\cR}\preceq\Sfunc$.
Then, for all $c\in\CSymbols$,
$\rarg{\cR}{\Sfunc}{c}=\emptyset$.
\end{fproposition}

\begin{proof}
We prove by contradiction that $\rarg{\cR}{\Seval_\cR}{c}=\emptyset$, and then 
the conclusion follows  by Theorem \ref{TheoRedundancyIsAntiMonotone}.
Let $t\in\GCTerms$ be such that $root(t)=c$. 
If $i\in\rarg{\cR}{\Seval_\cR}{c}$,
then $\Seval_\cR(t[s]_i)=\Seval_\cR(t[s']_i)$ for any $s,s'\in\GCTerms$ s.t. $s\neq s'$,
thus contradicting
$s \neq s'$.%
\end{proof}

\subsection*{Proofs of Section \ref{SecDecidability}}

\begin{lemma}\label{LemEquivDef}
For terms $t,t'$, and position $p\in\Pos(t)\cap\Pos(t')$,
the predicate 
$equiv(t,t',p) \equiv \exists s\in\GTerms.t'=t[s]_p$
is \emph{WS$k$S} definable.
\end{lemma}

\begin{proof*}
Assume that
the term $t$ is represented by $\sksT{T}$, and
the term $t'$ is represented by $\sksT{T'}$.
Then:
\[equiv(\sksT{T},\sksT{T'},p) \stackrel{def}{=}
\begin{array}[t]{@{}l@{}}
\forall q.( \neg(p \leq q) \Rightarrow
   \bigwedge_{f \in\Symbols} ( q \in T_f \Leftrightarrow q \in T'_f)) \hspace{.5cm}\proofbox
\end{array}\]
\end{proof*}

\begin{proposition}\label{PropAltRedArg}
Let
$\Sfunc$ be a term semantics for a signature $\Symbols$, $f\in\Symbols$, and
$i\in\{1,\ldots,ar(f)\}$.
The $i$-th argument of $f$ is {\em redundant w.r.t.\ $\Sfunc$\/} if
for all term $t\in\GTerms$, for all $p \in \Pos(t)$ such that $root(t|_p)=f$, and
for all $t'\in\GTerms$ such that
$equiv(t,t',p.i)$ is true,
$\Sfunc(t)=\Sfunc(t')$.
\end{proposition}

\begin{proof}
Immediate.
\end{proof}

\begin{ftheorem}{\ref{TheoDeciRed}}
Let $\Sfunc$ be a term semantics for a signature $\Symbols$.
If $\Sfunc$ is WS$k$S definable, then
redundancy w.r.t.\ $\Sfunc$\/
is decidable.
\end{ftheorem}

\begin{proof}
By using Proposition~\ref{PropAltRedArg}.
Assuming that $\Sfunc$ is defined in WS$k$S
by the formula $\Phi(\sksT{X},\sksT{Y})$,
redundancy is WS$k$S definable by using the following formula:
\[\begin{array}{c}
\forall \sksT{T}\;
\forall \sksT{S}\;
\forall \sksT{W}\;
\forall p \in T.(
Term(\sksT{T}) \wedge Term(\sksT{S}) \wedge Term(\sksT{W}) ~\wedge\\
p \in T_f \wedge equiv(\sksT{T},\sksT{S},p.i) ~\Rightarrow
(\Phi(\sksT{T},\sksT{W}) \Leftrightarrow
\Phi(\sksT{S},\sksT{W})))
\end{array}\]
Now, by Lemma \ref{LemEquivDef} and Theorem \ref{TheorRabin}, 
redundancy is decidable.
\end{proof}

\begin{fproposition}{\ref{TheoDecRedIsAntiMonotone}}
Let $\Sfunc,\Sfunc'$ be term semantics for
a signature $\Symbols$. If $\Sfunc \preceq \Sfunc'$, $\Sfunc'$ is WS$k$S definable,
and there exists a window set $T \subseteq \GTerms$
of $\Sfunc'$ w.r.t.\ $\Sfunc$ which
is WS$k$S definable,
then $\Sfunc$ is WS$k$S definable.
\end{fproposition}

\begin{proof}
Assuming
that semantics $\Sfunc'$ is defined by the WS$k$S formula
$\Phi'(\sksT{X},\sksT{Y})$, and that set $T$ is defined by the
WS$k$S formula $\Omega(\sksT{X})$, we build the following formula
defining $\Sfunc$: $\Phi(\sksT{X},\sksT{Y}) \stackrel{def}{=}
\Phi'(\sksT{X},\sksT{Y}) \wedge \Omega(\sksT{Y})$
\end{proof}

\begin{ftheorem}{\ref{TheoNHFDefinable}}
The set $\HNF_{\cR}$ of a finite left-linear, right-ground TRS $\cR$ is
WS$k$S definable.
\end{ftheorem}

\begin{proof}
Since the set $\REDEX_{\cR}$ of all redexes of a TRS $\cR$ is
WS$k$S definable \cite{Gallier-Book85}, and the set
$(\to^{\ast}_{\cR})[L]=\{t \in \GTerms \mid \exists s\in
L.t\to^{\ast}_{\cR} s\}$ is WS$k$S definable\footnote{Actually,
the set $(\to^{\ast}_{\cR})[L]$ is recognizable for every
recognizable tree language $L$~\cite{Com00}. Hence,
by \cite{TW68} it is WS$k$S definable.} for any regular set of terms $L$. We
can formulate the set $\HNF_{\cR}$ as:
$\Phi(\sksT{X}) = \neg \Omega(\sksT{X})$, where the set
$(\to^{\ast}_{\cR})[\REDEX_{\cR}]$ is defined by the predicate
$\Omega(\sksT{X})$.
\end{proof}

\begin{ftheorem}{\ref{CoroDecidRed}}
For a left-linear, right-ground TRS $\cR$ over a finite signature
$\Symbols$, the redundancy w.r.t.\ semantics
$\Sred_\cR$, $\Shnf_\cR$, $\Snf_\cR$, and $\Seval_\cR$
 is decidable.
\end{ftheorem}

\begin{proof}
Since the semantics $\Sred_{\cR}$ is WS$k$S definable, and the sets
$\NF_{\cR}$ and $\GCTerms$ are WS$k$S definable, by
Proposition \ref{TheoDecRedIsAntiMonotone}
we obtain that
the semantics $\Snf_{\cR}$ and $\Seval_{\cR}$ are WS$k$S definable.
Then, by considering also
Theorem \ref{TheoNHFDefinable} we obtain that
the semantics $\Shnf_{\cR}$ is WS$k$S definable.
Finally, by Theorem \ref{TheoDeciRed}, redundancy is decidable
for semantics $\Sred_\cR$, $\Shnf_\cR$, $\Snf_\cR$, and $\Seval_\cR$.
\end{proof}

\subsection*{Proofs of Section \ref{approxs}}

\begin{ftheorem}{\ref{TheoApproximationOfRedundancyOfEval}}
Let $\cR$ be a TRS, $\cR'$ be an approximation of $\cR$,
 $f\in\Symbols$, $i\in\{1,\ldots,ar(f)\}$,
and $\Sfunc\in\{\Seval,\Snf\}$. If $\cR$ is $\Sfunc_\cR$-defined
and $\Sfunc_{\cR'}$ is determined w.r.t.\ $f$ and $i$,
then $i\in\rarg{\cR}{\Sfunc_\cR}{f}$.
\end{ftheorem}

\begin{proof}
We prove the result for $\Seval_\cR$; the proof for $\Snf$ is analogous.
Note that, since $\cR$ is $\Seval_\cR$-defined,
$|\Seval_\cR(C[t])|\geq 1$ and $|\Seval_\cR(C[t[s]_i])|\geq 1$.
Moreover, since 
$\Seval_{\cR'}$ is determined w.r.t.\ $f$ and $i$,
$|\Seval_\cR(C[t])|=1$ and $|\Seval_\cR(C[t[s]_i])|=1$. Otherwise,
since $C[t[\Omega]_i]\to^*_{\cR'}C[t]$, 
$C[t[\Omega]_i]\to^*_{\cR'}C[t[s]_i]$,
$\to^*_\cR\subseteq\to^*_{\cR'}$, and the constructor symbols of
$\cR$ and $\cR'$ are identical (since $\NF_\cR=\NF_{\cR'}$ and $\Omega$
is a defined symbol), 
we would
also have $|\Seval_{\cR'}(C[t[\Omega]_i])|>1$.

Assume that $i\not\in\rarg{\cR}{\Seval_\cR}{f}$. Then, there exist
${C[\;]}$, $t\in\GTerms$ such that $root(t)=f$, and $s\in\GTerms$ such that
$\Seval_\cR(C[t])\neq\Seval_\cR(C[t[s]_i])$.
Then, since
$|\Seval_\cR(C[t])|=1$ and $|\Seval_\cR(C[t[s]_i])|=1$, it follows
that $\delta\in\Seval_\cR(C[t])$ and
$\delta'\in\Seval_\cR(C[t[s]_i])$ verify $\delta\neq\delta'$.
By reasoning as above, this would mean that
$|\Seval_{\cR'}(C[t[\Omega]_i])|>1$ thus leading to a contradiction.
\end{proof}

\begin{ftheorem}{\ref{TheoDecidRedViaDeterminism}}
Let $\Sfunc$ be a term semantics for a signature
$\Symbols\cup\{\Omega\}$.
If\/ $\Sfunc$ is WS$k$S definable, then
it is decidable whether
$\Sfunc$ is determined w.r.t.\ $f$ and $i$.
\end{ftheorem}
\begin{proof}
Assuming that $\Sfunc$ is defined in WS$k$S
by the formula $\Phi(\sksT{X},\sksT{Y})$,
the property is WS$k$S definable by using the following formula:
\[\begin{array}{c}
\forall \sksT{T}\;\forall p\;
Term(\sksT{T}) \wedge
p \in T_f \wedge p.i\in T_\Omega~\Rightarrow 
(\forall q\in T_\Omega, q=p.i)~\wedge \\
\forall\sksT{S}\;
 Term(\sksT{S}) \wedge \Phi(\sksT{T},\sksT{S}) \wedge \forall
\sksT{W}\;(Term(\sksT{W}) \wedge \Phi(\sksT{T},\sksT{W})\Rightarrow
\sksT{W}=\sksT{S})
\end{array}\]
Now, by Theorem \ref{TheorRabin}, the conclusion follows.
\end{proof}

\begin{ftheorem}{\ref{TheoDecidRedViaDeterminismForRG}}
Let $\cR$ be a left-linear TRS, $\cR_{rg}$ be the approximation $rg$ of $\cR$,
 $f\in\Symbols$, $i\in\{1,\ldots,ar(f)\}$,
and $\Sfunc\in\{\Seval_{\cR_{rg}},\Snf_{\cR_{rg}}\}$.
It is decidable whether
$\Sfunc$ is determined w.r.t.\ $f$ and $i$.
\end{ftheorem}

\begin{proof}
Since $\cR_{rg}$ is a left-linear and right ground TRS,
the semantics $\Sred_{\cR_{rg}}$ is WS$k$S definable.
The sets $\NF_{\cR}$ and $\GCTerms$ are WS$k$S definable and,
by
Proposition \ref{TheoDecRedIsAntiMonotone},
we obtain that
the semantics $\Snf_{\cR_{rg}}$ and $\Seval_{\cR_{rg}}$ are WS$k$S definable.
Then, by Theorem~\ref{TheoDecidRedViaDeterminism},
we obtain that the property
$\forall C[\:],t\in\GTerms, root(t)=f, i\in\{1,\ldots,ar(f)\}$,
$|\Sfunc(C[t[\Omega]_i])|\leq 1$
is decidable for
$\Sfunc\in\{\Seval_{\cR_{rg}},\Snf_{\cR_{rg}}\}$.
\end{proof}

\subsection*{Proofs of Section \ref{SecCharRed}}

\begin{ftheorem}{\ref{TheoRedundancyPositionIsAntiMonotone}}
Let $\Sfunc,\Sfunc'$ be term semantics
for a signature $\Symbols$. If $\Sfunc\preceq\Sfunc'$, then, for all
$t\in\Terms$, $\rpos{}{\Sfunc'}{t}\subseteq\rpos{}{\Sfunc}{t}$.
\end{ftheorem}

\begin{proof}
By contradiction.
Given $t\in\Terms$ and $p\in\rpos{\cR}{\Sfunc'}{t}$, by Definition \ref{DefStrongRedPos} 
we have that, 
for all $t',s \in \GTerms$
such that $t$ and $t'$ are $p$-prefix-equal,
$\Sfunc(t')=\Sfunc(t'[s]_p)$.
Now, since $\Sfunc\preceq\Sfunc'$,
there exists $T\subseteq\GTerms$ such that
$\Sfunc(t')=\Sfunc'(t')\cap T=\Sfunc'(t'[s]_p])\cap T=\Sfunc(t'[s]_p])$.
Hence, $p\in\rpos{\cR}{\Sfunc}{t}$.%
\end{proof}

\begin{fproposition}{\ref{PropNoRposCTerm}}
Let $\cR$ be a TRS such that $|\GCTerms|>1$, and
 $\Sfunc$ be a rewriting semantics such that $\Seval_{\cR}\preceq\Sfunc$.
Then, for all $t\in\CTerms$,
$\rpos{\cR}{\Sfunc}{t}=\emptyset$.
\end{fproposition}

\begin{proof}
We prove by contradiction that $\rpos{\cR}{\Seval_\cR}{t}=\emptyset$, and then 
the conclusion follows  by Theorem \ref{TheoRedundancyPositionIsAntiMonotone}.
Let $t\in\CTerms$.
If $p\in\rpos{\cR}{\Seval_{\cR}}{t}$,
then 
for all $t'\in\GTerms$ and $s,s'\in\GCTerms$ s.t. $t$ and $t'$ are $p$-prefix-equal
and $s\neq s'$,
$\Seval_{\cR}(t'[s]_p)=\Seval_{\cR}(t'[s']_p)$.
In concrete, for $t'\in\GCTerms$ s.t. $t'=\sigma(t)$ for some $\sigma\in\GCSubst$,
we should have $\Seval_{\cR}(t'[s]_p)=\{t'[s]_p\}=\{t'[s']_p\}=\Seval_{\cR}(t'[s']_p)$,
thus contradicting
$s\neq s'$.%
\end{proof}

\begin{fproposition}{\ref{PropRedArgAndRedPos}}
Let $\Sfunc$ be a term semantics for a signature $\Symbols$, $t\in\Terms$,
 $p\in\Pos(t)$, $f\in\DSymbols$.
For all positions $q,p'$ and $i\in\rarggen$ such that
$p=q.i.p'$ and $root(t|_q)=f$,
$p\in\rposgen$ holds.
\end{fproposition}

\begin{proof}
Let $t'=C[f(\ol{t})]_q$ such that $t$ and $t'$ are equal down to $p$.
Since $i\in\rarggen$, for all term
$s\in\GTerms$, $\Sfunc(C[f(\ol{t})]_q)=\Sfunc(C[f(\ol{t})[s]_i]_q)$.
In particular, if $s'=t'[s]_{q.i}|_p$, then
$\Sfunc(t')=\Sfunc(t'[s']_p)$ and the conclusion follows.
\end{proof}

\subsection*{Proofs of Section \ref{SecUsingRedPosCharRedVarCase}}

We recall here the notion of descendants of a position in a rewrite
sequence. This notion is usually meaningful for orthogonal TRS's
(since descendants of redexes, called residuals, are also redexes) but
it makes sense for arbitrary TRS's; we must always provide the
concrete rule applied at each rewriting step, see \cite[Section
4.2]{Terese03}.

\begin{definition}[\citeNP{HL91}]%
Let $A:t \stackrel{p}{\lto}_{l \to r} s$
and $q\in\Pos(t)$.
The set $q\bsh A$ (alternatively, $q\bsh\tuple{p,l{\to}r}$)
of descendants of $q$ in $s$ is defined as follows:
\[q\bsh A =\left\{\begin{array}{ll}
\{q\} & $if $ q < p $ or $ q \parallel p, \\
\{p.p_3.p_2 \mid r|_{p_3} = l|_{p_1}\} &
$if $ q=p.p_1.p_2 $ with $ 
    p_1\in\Pos_{\Variables}(l),\\
\emptyset & $otherwise.$
\end{array}\right.\]
\end{definition}
If $Q\subseteq\Pos(t)$ then $Q\bsh A$ denotes the set
$\bigcup_{q\in Q} q\bsh A$.
The notion of descendant extends to rewrite sequences in the obvious way.
If $Q$ is a set of pairwise disjoint positions in $t$ and $A:t \to^* s$,
then the positions in $Q\bsh A$ are pairwise disjoints.

\begin{fproposition}{\ref{PropRedAndVarInRhs}}
Let
$\cR$ be a left-linear CS,
$f \in \DSymbols$, and $i\in\{1,\ldots,ar(f)\}$.
Let $t\in\GTerms$,
$P\subseteq Pos_{f,i}(t)\cup\rpos{}{\Seval_\cR}{t}$
be a set of disjoint positions,
and $\ol{s}\in\GTerms$.
Let $t \to^* \delta$ for some $\delta\in\GCTerms$.
If,
for all $l\to r\in\cR_f$, $l|_i$ is a variable
which is $(f,i)$-redundant
in $r$,  then 
$t[\ol{s}]_P \to^* \delta$.
\end{fproposition}

\begin{proof*}
We prove by induction on $n$, $t \to^n \delta$. 
We assume $|\GCTerms|>1$, which is necessary for Proposition~\ref{PropNoRposCTerm} used below;
otherwise the proof is trivial.

\begin{enumerate}[2.]
\item
If $n=0$, then $t=\delta$, $Pos_{f,i}(t)=\emptyset$, and
by Proposition~\ref{PropNoRposCTerm}, $\rpos{}{\Seval_\cR}{t}=\emptyset$.
Hence, $P=\emptyset$ and $t[\ol{s}]_P = t = \delta$.

\item
If $n>0$, then
$t \stackrel{q}{\lto}_{l \to r} t' \to^{n-1} \delta$.
Let $P_i = P \cap Pos_{f,i}(t)$
and $P_{rpos} = P \setminus P_i$.
Now, we prove that
$t[\ol{s}|_{P_i}]_{P_i} \to^* \delta$.

\begin{itemize}
\item $(root(l)\neq f)$
Let 
$P_q = \{p \in P_i \mid q \leq p\}$
and 
$P'_i = P_i \setminus P_q$.
Then,
$P'_i\bsh\tuple{q,l{\to}r}=P'_i$.
By induction hypothesis,
$t[\ol{s}|_{P'_i}]_{P'_i} \stackrel{q}{\lto}_{l \to r} t'[\ol{s}|_{P'_i}]_{P'_i} \to^* \delta$
(if $\nexists p \in P'_i.p\leq q$),
or
$t[\ol{s}|_{P'_i}]_{P'_i} = t'[\ol{s}|_{P'_i}]_{P'_i} \to^* \delta$
(if $\exists p \in P'_i.p\leq q$).
Now, we prove 
$t[\ol{s}|_{P'_i\cup P_q}]_{P'_i\cup P_q} \to^* \delta$.

We have $P_q \bsh\tuple{q,l{\to}r} \subseteq Pos_{f,i}(t')$, 
since $\cR$ is a CS and this implies
each position in $P_q$ is under a variable of $l$,
i.e., $\forall q.p \in P_q$, $\exists p'\in\Pos_\Variables(l).p'\leq p$.
Then, $t[\ol{s}|_{P'_i \cup P_q}]_{P'_i \cup P_q} \stackrel{q}{\lto}_{l \to r} t''$ for some $t''\in\GTerms$,
since $\cR$ is left-linear.
Now $P'_i \cup (P_q \bsh\tuple{q,l{\to}r}) \subseteq Pos_{f,i}(t')$ and, by induction hyphotesis,
$t''=t'[\ol{w}]_{P'_i\cup(P_q \bsh\tuple{q,l{\to}r})} \to^* \delta$
for some $\ol{w}\in\GTerms$.

Hence,
$t[\ol{s}|_{P_i}]_{P_i} \to^* \delta$.

\item $(root(l) = f)$
Let 
$P_{q.i} = \{p \in P_i \mid q.i \leq p\}$
and 
$P'_i = P_i \setminus P_{q.i}$.
As in the previous case, we have
$t[\ol{s}|_{P'_i}]_{P'_i} \to^* \delta$.
Now, we prove 
$t[\ol{s}|_{P'_i\cup P_{q.i}}]_{P'_i\cup P_{q.i}} \to^* \delta$.

We have that $t[\ol{s}|_{P'_i\cup P_{q.i}}]_{P'_i\cup P_{q.i}} \stackrel{q}{\lto}_{l \to r} t''$ for some $t''\in\GTerms$,
since $l|_i$ is a variable, say $x$, and $\cR$ is left-linear.  Since
$x$ is $(f,i)$-redundant in $r$, $q.\Pos_x(r) \subseteq
Pos_{f,i}(t')\cup\rpos{}{\Seval_\cR}{t'}$.  
We also have that for all $p\in
P_{q.i}\bsh\tuple{q,l{\to}r}$, there is $p'\in q.\Pos_x(r)$ such that
$p' \leq p$.  
Hence, by induction hypothesis,
$t''=t'[\ol{w}]_{P'_i\cup P_{q.i}\bsh\tuple{q,l{\to}r}} \to^* \delta$
for some $\ol{w}\in\GTerms$.

Then,
$t[\ol{s}|_{P'_i \cup P_{q.i}}]_{P'_i \cup P_{q.i}} \to^* \delta$,
i.e.,
$t[\ol{s}|_{P_i}]_{P_i} \to^* \delta$.
\end{itemize}

Finally,
$(t[\ol{s}|_{P_i}]_{P_i})[\ol{s}|_{P_{rpos}}]_{P_{rpos}} \to^* \delta$ by definition.
Hence, $t[\ol{s}]_{P} \to^* \delta$.
\hfill\proofbox
\end{enumerate}
\end{proof*}

\begin{ftheorem}{\ref{TheorRedAndVarInRhs}}
Let $\cR$ be a left-linear CS. Let
$f \in \DSymbols$ and $i\in\{1,\ldots,ar(f)\}$.
If,
for all $l\to r\in\cR_f$, $l|_i$ is a variable
which is $(f,i)$-redundant
in $r$,  then $i\in\rarg{\cR}{\Seval_\cR}{f}$.
\end{ftheorem}

\begin{proof}
Let $C[\;]$ be a context such that $C|_p=\Box$,
and $t,s \in \GTerms$ be terms such that $root(t)=f$.
By Proposition~\ref{PropRedAndVarInRhs},
$\forall \delta\in\GCTerms$ s.t. $C[t] \to^* \delta$, $C[t[s]_i] \to^* \delta$
and viceversa.
Hence, 
$\Seval_{\cR}(C[t])=\Seval_{\cR}(C[t[s]_i])$.
\end{proof}

\subsection*{Proofs of Section \ref{SecUsingRedPosCharRedPatternCase}}

\begin{fproposition}{\ref{PropLemOrthoRedTupleGeneral}}
Let
$\cR$ be a left-linear, confluent, and $\Seval_\cR$-defined CS.
Let $f \in \DSymbols$ and $i\in\{1,\ldots,ar(f)\}$.
Let $t\in\GTerms$,
$P\subseteq Pos_{f,i}(t)\cup\rpos{}{\Seval_\cR}{t}$
be a set of disjoint positions,
and $a$ be a constant.
Let $t \to^* \delta$ for some $\delta\in\GCTerms$.
If
 $\cR$ is
$(f,i)$-joinable, then 
$t[\ol{a}]_P \to^* \delta$.
\end{fproposition}

\begin{proof*}
We prove by induction on $n$, $t \to^n \delta$.
We assume $|\GCTerms|>1$, which is necessary for Proposition~\ref{PropNoRposCTerm} used below;
otherwise the proof is trivial.

\begin{enumerate}[2.]
\item
If $n=0$, then $t=\delta$, $Pos_{f,i}(t)=\emptyset$, and
by Proposition~\ref{PropNoRposCTerm}, $\rpos{}{\Seval_\cR}{t}=\emptyset$.
Hence, $P=\emptyset$ and $t[\ol{a}]_P = t = \delta$.

\item
If $n>0$, then
$t \stackrel{q}{\lto}_{l \to r} t' \to^{n-1} \delta$.
Let $P_i = P \cap Pos_{f,i}(t)$
and $P_{rpos} = P \setminus P_i$.
Now, we prove that
$t[\ol{a}]_{P_i} \to^* \delta$.

\begin{itemize}
\item $(root(l)\neq f)$
Let 
$P_q = \{p \in P_i \mid q \leq p\}$
and 
$P'_i = P_i \setminus P_q$.
Then,
$P'_i\bsh\tuple{q,l{\to}r}=P'_i$.
By induction hypothesis,
$t[\ol{a}]_{P'_i} \stackrel{q}{\lto}_{l \to r} t'[\ol{a}]_{P'_i} \to^* \delta$
(if $\nexists p \in P'_i.p\leq q$), 
or
$t[\ol{a}]_{P'_i} = t'[\ol{a}]_{P'_i} \to^* \delta$
(if $\exists p \in P'_i.p\leq q$).
Now, we prove 
$t[\ol{a}]_{P'_i\cup P_q} \to^* \delta$.

We have $P_q \bsh\tuple{q,l{\to}r} \subseteq Pos_{f,i}(t')$, 
since $\cR$ is a CS and this implies
each position in $P_q$ is under a variable of $l$,
i.e., $\forall q.p \in P_q$, $\exists p'\in\Pos_\Variables(l).p'\leq p$.
Then, $t[\ol{a}]_{P'_i \cup P_q} \stackrel{q}{\lto}_{l \to r} t''$ for some $t''\in\GTerms$,
since $\cR$ is left-linear.
Now $P'_i \cup (P_q \bsh\tuple{q,l{\to}r}) \subseteq Pos_{f,i}(t')$ and by induction hyphotesis,
$t''=t'[\ol{a}]_{P'_i\cup(P_q \bsh\tuple{q,l{\to}r})} \to^* \delta$.

Hence,
$t[\ol{a}]_{P_i} \to^* \delta$.

\item $(root(l) = f)$
Since $P$ is a disjoint set, we have
$P'_i = P_i \setminus \{q.i\}$.
As in the previous case, we have
$t[\ol{a}]_{P'_i} \to^* \delta$.
Now, 
we prove 
$t[a]_{q.i} \to^* \delta$.

\begin{itemize}
\item $(l|_i\in\Variables)$
Let $l|_i=x$.
Then,
$t[a]_{q.i} \stackrel{q}{\lto}_{l \to r} t'[\ol{a}]_{q.Pos_x(r)}$.
Since $\cR$ is $(f,i)$-joinable,
$q.\Pos_x(r) \subseteq Pos_{f,i}(t')\cup\rpos{}{\Seval_\cR}{t'}$.
Thus, by induction hypothesis, 
$t'[\ol{a}]_{q.Pos_x(r)} \to^* \delta$.

\item $(l|_i\not\in\Variables)$
By $\Seval_\cR$-definedness,
there exist $l'\to r'\in\cR_f$ and $\sigma'\in\GSubst$
such that
$t[a]_{q.i}|_q \to^* \sigma'(l')$.
Assume $l'\to r'$ and $l\to r$ are different rules; otherwise $l|_i=t|_i=a$ and it is trivial.
Then,
$l$ and $l'$ unify up to the $i$-th argument
with mgu $\theta$. 
Thus,
$\langle l\to r,l'\to r',\theta\rangle$ is
a joinable 
$(f,i)$-triple  
of $\cR$.

Moreover,
we have that
there exist substitutions
$\tilde\sigma,\phi,\phi'\in\GSubst$
that split $\sigma$ and $\sigma'$ in terms of $\theta$,
i.e.,
such that
$\sigma(l) \to^* \tilde\sigma(l)$,
$\tilde\sigma=\phi\circ\theta$, 
and $\sigma'=\phi'\circ\theta$.
By
joinability 
of
$(f,i)$-triples, 
there exists $w\in\Terms$ such that
$\theta_\CSymbols(\tau_{l}(r)) \to^* w$ and $\theta_\CSymbols(\tau_{l'}(r')) \to^* w$.
By stability of $\to^*$,
$\phi\circ\theta_\CSymbols(\tau_{l}(r)) \to^* \phi(w)$ and
$\phi'\circ\theta_\CSymbols(\tau_{l'}(r')) \to^* \phi'(w)$.
By definition of $\theta_\CSymbols$ and left-linearity, 
$\phi(x)=\phi'(x)$ for
$x\not\in\Var(\theta(l)|_i) \cup \Var(\theta(l')|_i$,
and thus $\phi(w)=\phi'(w)$.
Summarizing, we have 
$t[\phi\circ\theta(r)]_q \to^* \delta$,
$t[\phi\circ\theta_\CSymbols(\tau_{l}(r))]_q \to^* t[\phi(w)]_q$,
$t[\phi'\circ\theta_\CSymbols(\tau_{l'}(r'))]_q \to^* t[\phi(w)]_q$,
and
$t[\phi'\circ\theta(r')]_q \to^* \delta'$ for some $\delta'\in\GCTerms$.
Now, we have to prove that $t[\phi(w)]_q \to^* \delta$ and $\delta'=\delta$.

Consider the set $P\subseteq\Pos(t')$ of positions
where $t[\phi\circ\theta(r)]_q$ and $t[\phi\circ\theta_\CSymbols(r)]_q$ differ.
By definition of $\tau_{l}$, $P \subseteq Pos_{f,i}(t')\cup\rpos{}{\Seval_\cR}{t'}$.
By induction hypothesis, 
$t[\phi\circ\theta_\CSymbols(\tau_{l}(r))]_q \to^* \delta$.
Thus, by confluence, $t[\phi\circ\theta_\CSymbols(\tau_{l}(r))]_q \to^* t[\phi(w)]_q \to^* \delta$.
And also $t[\phi'\circ\theta_\CSymbols(\tau_{l'}(r'))]_q \to^* t[\phi(w)]_q \to^* \delta$.

Finally, we prove that $\delta'=\delta$.
\begin{itemize}
\item
If $l'|_i\not\in\Variables$, then $l'|_i=a$, $\Var(l'|_i)=\emptyset$,
and
$\phi'\circ\theta_\CSymbols(\tau_{l'}(r'))=\phi'\circ\theta(r')$.
Thus, $\delta'=\delta$.
\item
If $l'|_i\in\Variables$, then $l'|_i=x$ and, by definition of $\tau_{l'}$,
$\phi'\circ\theta_\CSymbols(\tau_{l'}(r'))=\phi'\circ\theta_\CSymbols(r')$.
Since $\theta_\CSymbols(x)=a$, we have $\phi'\circ\theta_\CSymbols(r')=\phi'\circ\theta(r')$.
Thus $\delta'=\delta$.
\end{itemize}
\end{itemize}

Then,
$t[\ol{s}|_{P'_i \cup \{q.i\}}]_{P'_i \cup \{q.i\}} \to^* \delta$,
i.e.,
$t[\ol{s}|_{P_i}]_{P_i} \to^* \delta$.
\end{itemize}

Finally,
$(t[\ol{s}|_{P_i}]_{P_i})[\ol{s}|_{P_{rpos}}]_{P_{rpos}} \to^* \delta$ by definition.
Hence, $t[\ol{s}]_{P} \to^* \delta$.
\hfill\proofbox
\end{enumerate}
\end{proof*}

\begin{ftheorem}{\ref{CoroRedAndInfEvalComp}}
Let $\cR$ be
a left-linear, confluent
and
$\Seval_\cR$-defined CS. Let
$f \in \DSymbols$ and 
\linebreak
$i\in\{1,\ldots,ar(f)\}$.
If
 $\cR$ is
$(f,i)$-joinable, then $i\in\rarg{\cR}{\Seval_\cR}{f}$.
\end{ftheorem}

\begin{proof}
Let $a\in\CSymbols$ be a constant.
Let $C[\;]$ be a context such that $C|_p=\Box$,
and $t,s \in \GTerms$ be terms such that $root(t)=f$.
By Proposition~\ref{PropLemOrthoRedTupleGeneral},
we have that
$\forall \delta\in\GCTerms$ s.t. $C[t] \to^* \delta$, $C[t[c]_i] \to^* \delta$.
By confluence,
$\Seval_{\cR}(C[t])=\Seval_{\cR}(C[t[a]_i])$
and
$\Seval_{\cR}(C[t[s]_i])=\Seval_{\cR}(C[t[a]_i])$.
Hence,
$\Seval_{\cR}(C[t]) = \Seval_{\cR}(C[t[s]_i])$.
\end{proof}

\subsection*{Proofs of Section \ref{SecErasingRedArgs}}

\begin{proposition}\label{PropSustitucionFuncMuContraccion}
Let $\rho$ be a syntactic erasure for a signature $\Symbols$,
$t\in\Terms$ and $\sigma\in\Subst$.
Let $\sigma_\rho\in{{\it Subst}(\Symbols_\rho,\Variables)}$ be
such that $\sigma_\rho(x)=\tau_\rho(\sigma(x))$ for all $x\in\Variables$.
Then, $\tau_\rho(\sigma(t))=\sigma_\rho(\tau_\rho(t))$.
\end{proposition}

\begin{proof}
By structural induction.
If $t=x\in\Variables$, then the result is immediate,
since $\tau_\rho(x)=x$.
For the  induction step, we take $t=f(t_1,\ldots,t_n)$
for $t_1,\ldots,t_n\in\Terms$.
	\linebreak
Then,
$\tau_\rho(\sigma(t))=\tau_\rho(\sigma(f(t_1,\ldots,t_n)))=
\tau_\rho(f(\sigma(t_1),\ldots,\sigma(t_n)))=
	\linebreak
f_\rho(\tau_\rho(\sigma(t_{i_1})),\ldots,\tau_\rho(\sigma(t_{i_k})))$,
where
$\{1,\ldots,n\}-\rho(f)=\{i_1,\ldots,i_k\}$ and $i_m<i_{m+1}$ for $1\leq
m< k$.
By induction
hypothesis,
$f_\rho(\tau_\rho(\sigma(t_{i_1})),\ldots,\tau_\rho(\sigma(t_{i_k})))=
f_\rho(\sigma_\rho(\tau_\rho(t_{i_1})),\ldots,\sigma_\rho(\tau_\rho(t_{i_k})))$.
And finally,
$f_\rho(\sigma_\rho(\tau_\rho(t_{i_1})),\ldots,\sigma_\rho(\tau_\rho(t_{i_k})))=
	\linebreak
\sigma_\rho(f_\rho(\tau_\rho(t_{i_1}),\ldots,\tau_\rho(t_{i_k})))=
\sigma_\rho(\tau_\rho(t))$.
\end{proof}

\begin{lemma}\label{PropSemSetsOfRedArg}
Let $\Sfunc$ be a term semantics for a signature $\Symbols$. Let
$f\in\Symbols$, and $I\subseteq\rarggen$. Then, for all contexts
$C[\;]$ and for all $t,s_1,\ldots,s_k\in \GTerms$ such that
$root(t)=f$,
 $\Sfunc(C[t])=\Sfunc(C[t[\ol{s_k}]_I])$.
\end{lemma}

\begin{proof}
By induction on $k=|I|$. If $k=0$, it is immediate. If $k>0$, let
$i\in
I$ and $\ol{s'_{k-1}}=s_1,\ldots,s_{i-1},s_{i+1},\ldots,s_k$. By
the  induction hypothesis,
$\Sfunc(C[t])=\Sfunc(C[t[\ol{s'_{k-1}}]_{I-\{i\}}])$. Since $i$ is
redundant w.r.t.\ $\Sfunc$,
$\Sfunc(C[t[\ol{s'_{k-1}}]_{I-\{i\}}])=\Sfunc(C[t[\ol{s_k}]_I])$,
i.e.,
 $\Sfunc(C[t])=\Sfunc(C[t[\ol{s_k}]_I])$.
\end{proof}

\begin{lemma}\label{PropSemOfRedArgFromDifSymbols}
Let $\Sfunc$ be a term semantics for a signature $\Symbols$. Let
$f,g\in\Symbols$, and $i\in\rarggen$, $j\in\rarg{\cR}{\Sfunc}{g}$.
Then, for all contexts $C[\;]$ and for all $t,t',s,s'\in \GTerms$
such that $root(t)=f$ and $root(t')=g$,
$\Sfunc(C[t,t'])=\Sfunc(C[t[s]_i,t'[s']_j])$.
\end{lemma}

\begin{proof}
Let $C'[\:]$ be
the context $C'[\:]=C[\Box,t']$. By redundancy of $i$, we have
$\Sfunc(C[t,t'])=\Sfunc(C'[t])=\Sfunc(C'[t[s]_i])$. By redundancy of
$j$,
$\Sfunc(C'[t[s]_i])=\Sfunc(C[t[s]_i],t'])=
	\linebreak
\Sfunc(C[t[s]_i,t'[s']_j])$,
i.e.,
 $\Sfunc(C[t,t'])=\Sfunc(C[t[s]_i,t'[s']_j])$.
\end{proof}

Given a syntactic erasure $\rho$ and a term $t\in\GTerms$, we define
the
{\em maximal non-redundant context} $\MNRC^\rho(t)$ of $t$
as
$\MNRC^\rho(t)=t[\Box]_{p_1.\rho(f_1)}\cdots[\Box]_{p_n.\rho(f_n)}$,
where $p_1,\ldots,p_n$ are the positions of all outermost subterms
rooted
by symbols $f_1,\ldots,f_n$ such that $\rho(f_i)\neq\emptyset$ for
$1\leq i\leq n$. 

\begin{fproposition}{\ref{PropEquivErasurePreservesSemantics}}
If the syntactic erasure $\rho:\Symbols\to \pwset(\nat)$ is sound with
respect to the semantics $\Sfunc$, then for all $t,s\in\GTerms$,
$t\equiv_{\tau_\rho}s$ implies that $\Sfunc(t)=\Sfunc(s)$.
\end{fproposition}

\begin{proof}
By induction on the structure of $\MNRC^\rho(t)$ and using
Lemmata \ref{PropSemSetsOfRedArg} and
\ref{PropSemOfRedArgFromDifSymbols}.
\end{proof}

\begin{ftheorem}{\ref{TheoCompPreErasureEval}}
Let $\cR$ be a left-linear TRS,
$\Sfunc$ be a rewriting semantics for $\cR$
such that $\Sfunc \preceq \Sred_{\cR}$,
$\rho$ be a sound syntactic erasure for $\Sfunc$,
and $t,\delta\in\cT(\Symbols_\rho)$.
If $t\to^*_{\cR_\rho}\delta$, then
$\forall t',\delta'\in\GTerms$ such that $\tau_\rho(t')=t$ and
$\tau_\rho(\delta')=\delta$,
$\Sfunc(\delta')\subseteq\Sfunc(t')$.
\end{ftheorem}

\begin{proof}
By induction on the length $m$ of the derivation
$t\to^*_{\cR_\rho}\delta$.
If $m=0$, then  $t=\delta$, and
for all $t',\delta'\in\GTerms$ such that $\tau_\rho(t')=t$ and
$\tau_\rho(\delta')=\delta$,
by Proposition~\ref{PropEquivErasurePreservesSemantics},
$\Sfunc(t')=\Sfunc(\delta')$.

If $m>0$, then
$t\stackrel{\scriptsize p}{\to}_{\cR_\rho}s\to^*_{\cR_\rho}\delta$.
Consider $t'',\delta''\in\GTerms$ such that $\tau_\rho(t'')=t$ and
$\tau_\rho(\delta'')=\delta$.
First we prove, by induction on $p$,
that there exist $t',s'$ such that $\tau_\rho(t')=t$,
$\tau_\rho(s')=s$, and $t'\stackrel{\scriptsize p'}{\to}_\cR s'$.
\begin{enumerate}[2.]
\item
If $p=\toppos$, then
$t=\sigma(l)$ for some $l\to r$ in $\cR_\rho$.
Then, by Definition~\ref{DefPreErasureOfaTRS},
there exists $l'\to r'\in\cR$ such that $\tau_\rho(l')=l$ and
$\sigma_{l'}(\tau_\rho(r'))=r$.
Now,
there exist a term $t'\in\GTerms$ and a substitution $\sigma'$
such that $\tau_\rho(t')=t$ and $t'=\sigma'(l')$.
Then, by left-linearity,
for all $x$ in $\Var(l')$, if $x\in\Var(l)$, we have
$\sigma(x)=\tau_\rho(\sigma'(x))$.
Otherwise, let $l'|_q=x$,
$\sigma(x)=t'|_q$.
Hence, $t'\stackrel{\scriptsize \toppos}{\to}_\cR \sigma'(r')$.

\item
If $p=i.q$, then we consider the terms
$t=f_\rho(t_1,\ldots,t_k)$, $s=f_\rho(s_1,\ldots,s_k)$,
$t'=f(t'_1,\ldots,t'_n)$ and
$s'=f(s'_1,\ldots,s'_n)$,
such that $k=n - |\rho(f)|$.
Then, $p'=i'.q'$, where $i = | \{1 \leq i' \} - \rho(f)|$,
and for all $j,j'$ s.t. $1\leq j\leq k$ and $1\leq j'\leq n$,
$j = | \{1 \leq j' \} - \rho(f)|$,
$t_j=\tau_\rho(t'_{j'})$, $s_j=\tau_\rho(s'_{j'})$.
By the induction hypothesis, the conclusion follows.
\end{enumerate}

Now, by induction hypothesis,
for all $w,w'\in\GTerms$ such that $\tau_\rho(w)=s$ and
$\tau_\rho(w')=\delta$, we have that $\Sfunc(w')\subseteq\Sfunc(w)$.
Thus, since $w \equiv_{\tau_\rho} s'$ and
$w' \equiv_{\tau_\rho} \delta'$,
by Proposition~\ref{PropEquivErasurePreservesSemantics},
we have that $\Sfunc(\delta')\subseteq\Sfunc(s')$.
By definition of $\Sred$,
$\Sred_{\cR}(s') \subseteq \Sred_{\cR}(t')$.
Then, let $T$ be the window set such that $\Sfunc \preceq \Sred_{\cR}$,
$\Sred_{\cR}(s') \cap T \subseteq \Sred_{\cR}(t') \cap T $, and thus,
$\Sfunc(s') \subseteq \Sfunc(t')$.
Hence, we obtain that $\Sfunc(\delta') \subseteq \Sfunc(t')$.
But, by Proposition~\ref{PropEquivErasurePreservesSemantics},
$\Sfunc(t'')=\Sfunc(t')$ and $\Sfunc(\delta'')=\Sfunc(\delta')$;
thus, the conclusion follows.
\end{proof}

\begin{ftheorem}{\ref{TheoCorrPreErasure}}
Let $\cR$ be a left-linear TRS, $\Sfunc$ be a rewriting
semantics for $\cR$, $\rho$ be a sound
syntactic erasure for $\Sfunc$, and $t\in\GTerms$. If
$\delta\in\Sfunc(t)$, then
$\tau_\rho(t)\to^*_{\cR_\rho}\tau_\rho(\delta)$.
\end{ftheorem}

\begin{proof}
Let $t'=C[c,\ldots,c]$, where
$C[\Box,\ldots,\Box]=\MNRC^\rho(t)$ and $c\in\Symbols$ is the
constant used in $\cR_\rho$. Since $t\equiv_\rho t'$, by
Proposition \ref{PropEquivErasurePreservesSemantics}, $t\to^*_\cR\delta$
if and only if $t'\to^*_\cR\delta$. Now
we prove, by induction on the length $m$ of derivation $t'\to^*_\cR\delta$
that $\tau_\rho(t')\to^*_{\cR_\rho}\tau_\rho(\delta)$. If $m=0$, then
$t'=\delta$ and the result is
immediate. If $m>0$, we let $t'\stackrel{p}{\to}_{\cR}s\to^*_\cR\delta$.
By induction on $p$, we prove that either
$\tau_\rho(t')\to_{\cR_\rho}\tau_\rho(s)$ or $\tau_\rho(t')=\tau_\rho(s)$.
\begin{enumerate}[2.]
\item If $p=\toppos$, then there exists $l\to r$ in $\cR$ such that
$t'=\sigma(l)$ and $s=\sigma(r)$. By Proposition
\ref{PropSustitucionFuncMuContraccion},
$\tau_\rho(\sigma(l))=\sigma_\rho(\tau_\rho(l))$ and
$\tau_\rho(\sigma(r))=\sigma_\rho(\tau_\rho(r))$ where
$\sigma_\rho(x)=\tau_\rho(\sigma(x))$ for all $x\in\Variables$.
Left-linearity of $\cR$ ensures
that, every variable $x$ that occurs within an erasable subterm of
$l$ (i.e., a subterm $l|_p$ such that there exists $q.i<p$ such that
$i\in\rho(root(l|_q))$) does not occur in $\tau_\rho(l)$. Thus, when
considering $x\in\Var(l)-\Var(\tau_\rho(l))$,
by definition of $t'$, it must be $\sigma(x)=c$. Hence,
$\sigma_\rho(\tau_\rho(r))=\sigma_\rho(\sigma_l(\tau_\rho(r)))$ where
$\sigma_l$ is fixed as in Definition \ref{DefErasureOfaTRS}. Thus,
by definition of $\cR_\rho$, $\tau_\rho(t')\to_{\cR_\rho}\tau_\rho(s)$.
\item If $p=i.q$, then we let $t'=f(t'_1,\ldots,t'_k)$ and
$s=f(s_1,\ldots,s_k)$ and consider two cases:
\begin{enumerate}
\item If $i\in\rho(f)$, then $\tau_\rho(t')=\tau_\rho(s)$ since $t$ only
differs from $s$ in the $i$-th argument $t_i$ of $f$ in $t'$
(which is removed by $\tau_\rho$).
\item If $i\not\in\rho(f)$, then, the $i$-th argument of $f$ in $t$
becomes the (transformed) $j$-th argument $\tau_\rho(t_i)$ of
$f$ in $\tau_\rho(t')$, where $j=|\{1\leq i\}-\rho(f)|$.
By the induction hypothesis,
either $\tau_\rho(t'_i)\to_{\cR_\rho}\tau_\rho(s_i)$ or
$\tau_\rho(t'_i)=\tau_\rho(s_i)$. In both cases, the conclusion
follows.
\end{enumerate}
\end{enumerate}
Therefore, we have that either
$\tau_\rho(t')\to\tau_\rho(s)$ or $\tau_\rho(t')=\tau_\rho(s)$. By
the induction hypothesis,
$\tau_\rho(s)\to^*_{\cR_\rho}\tau_\rho(\delta)$.
Thus, $\tau_\rho(t')\to^*_{\cR_\rho}\tau_\rho(\delta)$. Since
$\tau_\rho(t)=\tau_\rho(t')$, the conclusion follows.
\end{proof}

\begin{ftheorem}{\ref{TheoCorrPreErasureEval}}
Let $\cR$ be a left-linear TRS,
 $\rho$ be a sound syntactic erasure for $\Seval_\cR$,
 $t\in\GTerms$, and $\delta\in\GCTerms$. Then,
$\tau_\rho(t)\to^*_{\cR_\rho}\delta$ iff $\delta\in\Seval_\cR(t)$.
\end{ftheorem}

\begin{proof}
Immediate from Theorem \ref{TheoCompPreErasureEval} and
Theorem \ref{TheoCorrPreErasure}.
\end{proof}

\begin{ftheorem}{\ref{TheoConfluentRho}}
Let $\cR$ be a left-linear TRS.
Let $\rho$ be a sound syntactic erasure for $\Seval_\cR$.
If $\cR$ is
$\Seval_\cR$-defined
and confluent, then the erasure
$\cR_\rho$ of $\cR$ is confluent.
\end{ftheorem}

\begin{proof}
Given $t\in\cT(\Symbols_\rho)$, if
$t_1 ~{}^{\:\:*}_{\cR_\rho}\!\!\!\!\leftarrow t \to^*_{\cR_\rho} t_2$ with
$t_1\neq t_2$,
by Theorem~\ref{TheoCompPreErasureEval}, there exist
$s,s_1,s_2\in\GTerms$ such that $\tau_\rho(s)=t$, $\tau_\rho(s_1)=t_1$,
$\tau_\rho(s_2)=t_2$,
$\Seval_\cR(s_1)\subseteq\Seval_\cR(s)$, and
$\Seval_\cR(s_2)\subseteq\Seval_\cR(s)$.

Since $\Seval_\cR$ is $\cR$-normalized, and $\cR$ is confluent and $\Seval_\cR$-defined,
$\Seval_\cR(s)$ is a singleton consisting of the normal form $t'$.
Moreover,
$\Seval_\cR(s_1)=\Seval_\cR(s_2)=\Seval_\cR(s)$, and by
Theorem~\ref{TheoCorrPreErasure},
$t_1 \to^*_{\cR_\rho} \tau_\rho(t') ~{}^{\:\:*}_{\cR_\rho}\!\!\!\!\leftarrow t_2$.
\end{proof}

\begin{ftheorem}{\ref{TheoWeakTermRho}}
Let $\cR$ be a left-linear and completely defined TRS, and
$\rho$ be a sound syntactic erasure for $\Seval_\cR$.
If $\cR$ is normalizing, then the erasure $\cR_\rho$ of $\cR$ is normalizing.
\end{ftheorem}

\begin{proof}
Since $\cR$ is normalizing and completely defined,
$\forall t\in\GTerms$, $\exists\delta\in\GCTerms\in\Seval_\cR(t)$.
Then, by Theorem~\ref{TheoCorrPreErasure},
$\tau_\rho(t) \to^*_{\cR_\rho} \tau_\rho(\delta)$, and,
by Proposition~\ref{PropThereIsNoRedArgForConstructors},
$\tau_\rho(\delta)=\delta$.
Hence, the conclusion follows.
\end{proof}

\section{Benchmarks Code}\label{AppCharRedPos}

We give some example programs 
which contain redundant arguments, borrowed from the literature 
and/or obtained by applying common transformation processes. 
For each example, we show 
the final program which results from optimizing the program
by using our automatic redundant argument removal prototype.
Programs are given in the (currified) functional programming syntax used by \Curry.

\subsection*{Program \pr{bogus}}

The following program \pr{bogus} is borrowed from \cite{Kobayashi00,WS99}, where it is introduced for 
useless variable elimination (UVE), a popular technique for removing dead variables.

\startprog
data Nat = Z | S Nat\\[-.2cm]

loop :: Nat -> Nat -> Nat -> Nat
loop a bogus Z     = loop (S a) (S bogus) (S Z)\nopagebreak
loop a bogus (S x) = a
\stopprog
The second argument of \pr{loop} is signaled as redundant and then removed.
\startprog
loop' :: Nat -> Nat -> Nat
loop' a Z     = loop' (S a) (S Z)\nopagebreak
loop' a (S x) = a
\stopprog

\subsection*{Program \pr{applast}}

The following program is borrowed from 
\cite{Leu98} and is obtained by program specialization \cite{AFV98}.
\sstartprog
data Nat = 0 | S Nat\\[-.2cm]

append::[Nat] -> [Nat] -> [Nat]     last::[Nat] -> Nat
append nil    y = y                 last (x:nil)  = x\nopagebreak
append (x:xs) y = x:(append xs y)   last (x:y:ys) = last (y:ys)
\sstopprog
The specialization of the program \pr{applast}
for goal `\pr{last (append xs (x:nil))}' yields:
\sstartprog
applast::[Nat] -> Nat -> Nat        lastnew::Nat -> [Nat] -> Nat -> Nat
applast nil    z = z                lastnew x nil    z = z\nopagebreak
applast (x:xs) z = lastnew x xs z   lastnew x (y:ys) z = lastnew y ys z
\sstopprog
The first argument of \pr{applast} and the first and second arguments of \pr{lastnew} 
are identified as redundant and removed.
\sstartprog
applast' :: Nat -> Nat              lastnew' :: Nat -> Nat
applast' z = z                      lastnew' z = z
\sstopprog

\subsection*{Program \pr{plus\_minus}}

This example is borrowed from 
\cite{Leu98} and is obtained by program specialization \cite{AFV98}.
\startprog
data Nat = Z | S Nat\\[-.2cm]

plus :: Nat -> Nat -> Nat      minus :: Nat -> Nat -> Nat
plus Z     x = x               minus x     Z     = x\nopagebreak
plus (S x) y = S (plus x y)    minus (S x) (S y) = minus x y
\stopprog
The specialization for goal `\pr{minus (plus x y) x}' yields:
\startprog
minus\_pe :: Nat -> Nat -> Nat
minus\_pe Z     y = y\nopagebreak
minus\_pe (S x) y = minus\_pe x y
\stopprog
The first argument of \pr{minus\_pe} is identified as redundant and removed.
\startprog
minus\_pe' :: Nat -> Nat
minus\_pe' y = y
\stopprog

\subsection*{Program \pr{plus\_leq}}

This example is borrowed from 
\cite{Leu98} and is obtained by program specialization \cite{AFV98}.
\startprog
data Nat = Z | S Nat\\[-.2cm]

plus :: Nat -> Nat -> Nat      leq :: Nat -> Nat -> Bool
plus Z     x = x               leq Z     x     = True\nopagebreak
plus (S x) y = S (plus x y)    leq (S x) Z     = False\nopagebreak
                               leq (S x) (S y) = leq x y
\stopprog
The specialization for goal `\pr{leq x (plus x y)}' yields:
\startprog
leq\_pe :: Nat -> Nat -> Bool
leq\_pe Z     x = True\nopagebreak
leq\_pe (S x) y = leq\_pe x y
\stopprog
Both arguments of \pr{leq\_pe} are identified as redundant and removed.
\startprog
leq\_pe' :: Bool
leq\_pe' = True
\stopprog

\subsection*{Program \pr{double\_even}}

This example is borrowed from \cite{Leu98} 
and is obtained by program specialization \cite{AFV98}.
\startprog
data Nat = Z | S Nat\\[-.2cm]

double :: Nat -> Nat               even :: Nat -> Bool
double Z     = Z                   even Z         = True\nopagebreak
double (S x) = S (S (double x))    even (S Z)     = False\nopagebreak
                                   even (S (S x)) = even x
\stopprog
The specialization for goal `\pr{even (double x)}' yields:
\startprog
even\_pe :: Nat -> Bool
even\_pe Z = True\nopagebreak
even\_pe (S x) = even\_pe x
\stopprog
The argument of \pr{even\_pe} is identified as redundant and removed.
\startprog
even\_pe' :: Bool
even\_pe' = True
\stopprog

\subsection*{Program \pr{sum\_allzeros}}

This example is borrowed from \cite{Leu98} 
and is obtained by program specialization \cite{AFV98}.
\startprog
data Nat = Z | S Nat\\[-.2cm]

plus :: Nat -> Nat -> Nat     sum :: [Nat] -> Nat\nopagebreak
plus Z     x = x              sum nil    = Z\nopagebreak
plus (S x) y = S (plus x y)   sum (x:xs) = plus x (sum xs)\\[-.2cm]

allzeros :: [Nat] -> [Nat]
allzeros nil    = nil\nopagebreak
allzeros (x:xs) = Z:(allzeros xs)
\stopprog
The specialization for goal `\pr{sum (allzeros x)}' yields:
\startprog
sum\_pe :: [Nat] -> Nat
sum\_pe nil    = Z\nopagebreak
sum\_pe (x:xs) = sum\_pe xs
\stopprog
The argument of \pr{sum\_pe} is identified as redundant and removed.
\startprog
sum\_pe' :: Nat
sum\_pe' = Z
\stopprog

\subsection*{Mutual Recursion 1}

This program is taken from Example 2.23 of \cite{AGTR01}.

\startprog
data Nat = Z | S Nat
f :: Nat -> Nat -> Nat
f Z     y = Z\nopagebreak
f (S x) y = f (f x y) y
\stopprog
Both arguments of \pr{f} are identified as redundant and removed.
\startprog
f' :: Nat
f' = Z
\stopprog

\subsection*{Mutual Recursion 2}

This program is taken from Example 2.24 of \cite{AGTR01}.

\startprog
data Nat = Z | S Nat
f :: Nat -> Nat
f Z         = S Z\nopagebreak
f (S Z)     = S Z\nopagebreak
f (S (S x)) = f (f (S x))
\stopprog
The argument of \pr{f} is identified as redundant and removed.
\startprog
f' :: Nat
f' = S Z
\stopprog

\label{lastpage}
\end{document}